\documentclass{article}

\usepackage{arxiv}

\usepackage[utf8]{inputenc} 
\usepackage[T1]{fontenc}    
\usepackage{hyperref}       
\usepackage{url}            
\usepackage{booktabs}       
\usepackage{amsfonts}       
\usepackage{nicefrac}       
\usepackage{microtype}      
\usepackage{lipsum}
\usepackage{graphicx}
\usepackage{tikz}
\usetikzlibrary{positioning}
\usepackage{amssymb}
\usepackage{xcolor}
\usepackage{array}
\usepackage{verbatim}
\usepackage{booktabs}
\usepackage{algorithmic}
\usepackage{url}
\usepackage{graphicx}
\usepackage{textcomp}
\usepackage{multirow}
\usepackage{pdfpages}
\usepackage{xcolor}
\usepackage{tikz}
\usepackage{setspace}
\usepackage{float}
\usepackage{tabularx}
\usepackage{makecell}
\usepackage{adjustbox}
\graphicspath{ {./images/} }

\title{SHIELDA: Structured Handling of Exceptions in LLM-Driven Agentic Workflows}

\author{
Jingwen Zhou \\
CSIRO's Data61 \\
Melbourne, Victoria, Australia \\
\texttt{helen.zhou@data61.csiro.au} \\
\And
Jieshan Chen \\
CSIRO's Data61 \\
Sydney, New South Wales, Australia \\
\texttt{jieshan.chen@data61.csiro.au} \\
\And
Qinghua Lu \\
CSIRO's Data61 \\
University of New South Wales \\
Sydney, New South Wales, Australia \\
\texttt{qinghua.lu@data61.csiro.au} \\
\And
Dehai Zhao \\
CSIRO's Data61 \\
Sydney, New South Wales, Australia \\
\texttt{dehai.zhao@data61.csiro.au} \\
\And
Liming Zhu \\
CSIRO's Data61 \\
University of New South Wales \\
Sydney, New South Wales, Australia \\
\texttt{liming.zhu@data61.csiro.au}
}

\begin{document}
\maketitle
\begin{abstract}
Large Language Model (LLM) agentic systems are software systems powered by LLMs that autonomously reason, plan, and execute multi-step workflows to achieve human goals, rather than merely executing predefined steps. During execution, these workflows frequently encounter exceptions. Existing exception handling solutions often treat exceptions superficially, failing to trace execution-phase exceptions to their reasoning-phase root causes. Furthermore, their recovery logic is brittle, lacking structured escalation pathways when initial attempts fail. To tackle these challenges, we first present a comprehensive taxonomy of 36 exception types across 12 agent artifacts. Building on this, we propose SHIELDA (\textbf{S}tructured \textbf{H}andl\textbf{I}ng of \textbf{E}xceptions in \textbf{L}LM-\textbf{D}riven \textbf{A}gentic Workflows), a modular runtime exception handling framework for LLM agentic workflows. SHIELDA uses an exception classifier to select a predefined exception handling pattern from a handling pattern registry. These patterns are then executed via a structured handling executor, comprising local handling, flow control, and state recovery, to enable phase-aware recovery by linking exceptions to their root causes and facilitate composable strategies. We validate SHIELDA’s effectiveness through a case study on the AutoPR agent, demonstrating effective, cross-phase recovery from a reasoning-induced exception. 
\end{abstract}


\section{Introduction}
\label{sec:intro}
The landscape of artificial intelligence (AI) is being reshaped by the rise of LLM-based agentic systems, which are software systems powered by LLMs that autonomously reason, plan and execute multi-step workflows to achieve human goals, rather than merely executing predefined steps.  During execution, these workflows frequently encounter exceptions which may propagate across different workflow phases. For instance, flawed reasoning logic, or hallucinated steps during planning can cascade into execution failures \cite{mirzadeh2024gsm}. Similarly, inconsistencies like overconfidence in hallucinated information or conflicting goals due to ambiguous task definitions can significantly derail an agent's decision-making \cite{xiong2023can}. The \textit{execution phase} concurrently presents its own set of hurdles. Dynamic tool invocation, where agents actively call external tools' APIs, frequently results in failures like API timeouts or cross-agent coordination issues \cite{ruan2023tptu}. The inherent iterative loop between these two core phases further demands continuous monitoring to prevent cascading failures across cycles. While existing works have addressed specific exception types within these phases \cite{chen2025towards, hou2025model}, the current understanding of exceptions in agentic workflows remains fragmented and isolated, lacking a comprehensive characterization of their scope and distribution. This highlights a critical need for a systematic understanding of the unique exceptions in agentic workflows.

The inherent dynamism of agentic workflows presents a fundamental challenge to conventional exception handling paradigms. Traditionally, exception handling in software systems has been rooted in deterministic, linear processes, proving inadequate for the unpredictable nature of autonomous agents \cite{jin2024llms}. For instance, static mechanisms like try-catch blocks are simply not equipped to deeply probe an agent's intricate decision-making during the reasoning/planning phase or to adapt to the fluid complexities of dynamic tool invocation and multi-agent coordination during execution \cite{hou2025model, kong2024sharpunlockinginteractivehallucination}. While the community has seen efforts to mitigate specific LLM-related exceptions, such as hallucination detection, these approaches often remain isolated. They typically fall short of offering a unified framework for comprehensive exception management that spans diverse agent modalities and workflow phases. Furthermore, the challenge extends beyond runtime handling to post-hoc analysis; recent work highlights that the process of failure attribution—identifying which agent and at which step a task failure originates—remains a complex, labor-intensive, and largely underexplored problem~\cite{zhang2025agent}. Our analysis reveals two critical and pervasive gaps in current solutions: a notable absence of phase-aware, modality-generalized exception handling and lack of support for structured escalation pathways. Addressing these profound shortcomings thus necessitates a systematic approach to organizing an effective exception handling mechanism and developing a truly unified framework.

\textcolor{black}{To address these challenges, we conducted a systematic analysis of real-world agentic workflow exceptions. Our observations revealed that exceptions often stem from two distinct sources: 1) the agent’s internal reasoning and planning processes; and 2) its execution actions. This insight, coupled with the cyclical nature of agentic workflows, guided our categorization of exceptions into Reasoning/Planning and Execution phases, providing a structured framework to disentangle their interconnected effects. Building upon this foundation, we propose \textbf{SHIELDA} (\textbf{S}tructured \textbf{H}andl\textbf{I}ng of \textbf{E}xceptions in \textbf{L}LM-\textbf{D}riven \textbf{A}gentic Workflows), a novel, modular runtime framework designed to systematically detect, classify, and handle these critical exceptions. This study is guided by three core research questions: What types of exceptions occur in agentic workflows, and how are they distributed across agentic workflow phases? What exception handling mechanisms have been proposed for LLM-based agentic systems, and how can they be systematically organized into a structured module of exception handling mechanisms? How effective is the proposed exception handling framework in recovering from agentic workflow exceptions at runtime?}

\textcolor{black}{Generally, we make the following contributions:}
\begin{itemize}
    \item \textbf{A Comprehensive Exception Taxonomy:} \textcolor{black}{Grounded in a systematic literature review of 55 studies, we construct a fine-grained taxonomy of 36 exception types across 12 agent artifacts. This taxonomy distinguishes exceptions across reasoning, planning, and execution phases, revealing two critical gaps in existing solutions: a lack of phase-aware handling and limited support for structured escalation.}
    \item \textbf{A Structured Exception Handling Approach:} We define a structured, triadic design for exception handling that organizes existing mechanisms into three orthogonal dimensions: local handling, flow control, and state recovery. This composable design serves as the foundation for creating reusable handler patterns within our SHIELDA framework.
    \item \textbf{The SHIELDA Framework:} We propose SHIELDA, a modular runtime framework that leverages our taxonomy and structured handling approach.
    \item \textbf{Empirical Validation:} \textcolor{black}{We provide an in-depth case study of the AutoPR agent to demonstrate SHIELDA's effectiveness. Our results reveal that SHIELDA can successfully handle the cross-phase exceptions.}
\end{itemize}

\section{Methodology}
\label{sec:method}
\begin{figure}
    \centering
    \includegraphics[width=1\linewidth]{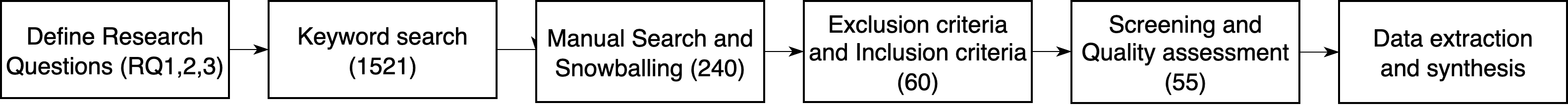}
    \caption{\textcolor{black}{Overview of the systematic literature review (SLR) process forming the core of our methodology.}}
    \label{fig:metho}
    \vspace{-10pt}
\end{figure}

\textcolor{black}{This section describes the methodology used to investigate exception types and handling mechanisms in agentic workflows. Our process includes five stages: (1) defining research questions, (2) conducting a systematic literature review (SLR), including screening, quality assessment, and data synthesis, (3) developing an exception taxonomy, (4) mapping existing handling strategies, and (5) using these insights to inform framework design. Figure~\ref{fig:metho} illustrates the process of our systematic literature review (SLR), which forms the empirical foundation of our methodology.}

\subsection{Research Questions and Scope}

\textcolor{black}{To guide our study, we formulated four research questions:}

\begin{itemize}
\item \textcolor{black}{RQ1: What types of exceptions occur in agentic workflows, and how are they distributed across agentic workflow phases?} \\
This question aims to characterize the exceptions that disrupt agentic workflows. Answering it requires the construction of a phase-aware taxonomy of exceptions grounded in real-world agent interactions, which forms the foundation for our subsequent analysis.

\item \textcolor{black}{RQ2: What exception handling mechanisms have been proposed for LLM-based agentic systems, and how can they be systematically organized into a structured model?} \\
This question seeks to move beyond ad-hoc mechanisms by systematically deconstructing existing handling mechanisms.

\item \textcolor{black}{RQ3: How can a framework based on our structured model be designed to handle real-world exceptions, and how effective is it at recovering from agentic workflow exceptions?} \\
This final, two-part question guides the design and validation of our SHIELDA framework. First, we propose a modular architecture that operationalizes the triadic model from RQ2. Second, we demonstrate its practical effectiveness through a case study, evaluating its ability to resolve a complex, cross-phase exception at runtime.

\end{itemize}

\subsection{Systematic Literature Review (SLR)}
To answer RQ1 and RQ2, we conducted a systematic literature review following standard software engineering practices. Our goal was to capture both exception types and the mechanisms used to handle them in the context of agentic workflows. We initiated the SLR with a structured keyword-based search across major academic databases including IEEE Xplore, ACM Digital Library, SpringerLink, and Google Scholar. To ensure comprehensive coverage, keywords were grouped into five semantic categories:

\begin{itemize}
    \item \textbf{Core Concepts:} LLM agent workflow, agentic workflow, agentic system, exception handling, error handling, failure management
    \item \textbf{Reasoning/Planning Errors:} reasoning exceptions, hallucination in LLMs, conflicting goals, ambiguous task definitions, reasoning trace validation
    \item \textbf{Execution Failures:} tool invocation failures, API failures, cross-agent coordination, server selection in workflows, MCP server crashes
    \item \textbf{Workflow Structures:} agentic workflows, multi-agent systems, workflow management, iterative workflows, cyclical workflows
    \item \textbf{AI/SE Context:} AI agent exceptions, software engineering exception handling, fault tolerance in AI systems
\end{itemize}

\textbf{Keyword search, manual search, and snowballing:} Boolean combinations (e.g., ``LLM agent'' AND ``exception handling'' AND ``tool failure'') yielded \textcolor{black}{1521} initial studies. To ensure domain relevance and improve coverage, we manually reviewed proceedings from top-tier venues. Forward and backward snowballing from anchor papers added \textcolor{black}{240} studies, resulting in \textcolor{black}{1761} candidates.

\textbf{Inclusion and Exclusion Criteria:}  
We defined filtering rules to ensure conceptual alignment and technical quality.

\begin{itemize}
    \item \textbf{Inclusion:}
    \begin{itemize}
        \item \textcolor{black}{Papers published in the last 10 years in English.}
        \item \textcolor{black}{Peer-reviewed studies or high-quality preprints (assessed by authors) with demonstrable rigor.}
        \item Studies on exceptions, recovery, or failure handling in LLM agents or multi-agent systems.
        \item Work discussing reasoning/planning errors (e.g., hallucination, faulty logic) or execution-level failures (e.g., tool crashes, coordination failures).
    \end{itemize}
    \item \textbf{Exclusion:}
    \begin{itemize}
        \item Work focused solely on LLM training or tuning, without runtime workflow concerns.
        \item General SE papers without relevance to LLMs, agents, or workflows.
        \item Abstract-only papers, vision pieces, or articles without empirical evidence.
        \item Failures unrelated to agent logic (e.g., hardware crashes or network layer issues).
    \end{itemize}
\end{itemize}

After filtering, \textcolor{black}{60} papers were retained.

\textbf{Screening and Quality Assessment:}  
We conducted a two-stage assessment. First, titles and abstracts were reviewed by two researchers for relevance, with discrepancies resolved via discussion. Second, full texts were assessed on a 5-point scale covering methodological clarity, empirical grounding, and relevance. A score of 3 or above was required for inclusion. This retained \textcolor{black}{55} high-quality papers.

\textbf{Data Extraction and Synthesis:}  
\textcolor{black}{For each selected study, we extracted publication metadata (Title, Author, Year, Source, Topic, etc.), related exception types, associated workflow phases, and exception handling mechanisms or strategies (\footnote{https://github.com/submissionpurposeonly/dataextraction}). All entries were compiled into a structured evidence matrix aligned to our research questions. Synthesis revealed two major phases where exceptions typically occur:}

\begin{itemize}
    \item \textbf{Reasoning and Planning (RP)} \textcolor{black}{phase refers to the stage where the agent analyzes its input, formulates goals, interprets context, and generates a task plan or decision strategy. These exceptions reflect cognitive-level breakdowns in the agent’s internal decision-making process before it acts on the world.}
    \item \textbf{Execution (E)} \textcolor{black}{phase encompasses the stage where the agent operationalizes its plan by invoking tools, calling APIs, generating outputs, or interacting with UIs and other systems. These exceptions represent operational breakdowns that occur during the execution of decisions or interaction with external components.}
\end{itemize}

This phase-oriented structure served as the foundation for the exception taxonomy (Table~\ref{tab:exception_taxonomy}) and informed the modular design of the SHIELDA framework introduced in Section~\ref{sec:exception_framework}.

\section{Exception Categorization and Handling Strategies}
\label{sec:exception_framework}
\textcolor{black}{This section addresses the first two research questions (RQ1 and RQ2) by constructing a taxonomy of exceptions in agentic workflows and analyzing the effectiveness of existing handling strategies. To answer RQ1, we identified and summarized 36 exception artifacts from the systematic literature review, covering exceptions across the reasoning/planning and execution phases. To address RQ2, we evaluate the limitations of current strategies, highlighting gaps in phase coverage, composability, and escalation support.}

\begin{table*}[ht]
\renewcommand{\arraystretch}{1.05}
\caption{A Taxonomy of Exceptions in Agentic Workflow Phases and Artifacts.}
\label{tab:exception_taxonomy}
\centering
\scriptsize

\begin{minipage}[t]{0.49\textwidth}
\centering
\begin{tabular}{p{1.8cm} p{4.2cm} p{0.5cm}}
\toprule
\textbf{Artifacts} & \textbf{Detailed Exceptions} & \textbf{Phases} \\
\midrule
Goal & Ambiguous Goal~\cite{chin2024human} & RP \\
& Conflicting Goal~\cite{guan2024intelligent} & RP \\
\midrule
Context & Context Corruption~\cite{zhan2024injecagent} & RP \\
& Context Ambiguity~\cite{chin2024human} & RP \\
\midrule
Reasoning & Contradictory Reasoning~\cite{sun2024ai} & RP \\
& Circular or Invalid Reasoning~\cite{yao2023react} & RP \\
\midrule
Planning & Faulty Task Structuring~\cite{xie2024travelplanner} & RP \\
& Overextended Planning & RP \\
\midrule
Memory & Memory Poisoning~\cite{chen2025agentpoison} & RP/E \\
& Outdated Memory~\cite{hatalis2023memory} & RP/E \\
& Misaligned Memory Recall~\cite{salama2025meminsightautonomousmemoryaugmentation} & RP/E \\
\midrule
Knowledge Base & Hallucinated Facts~\cite{zong2024triad} & RP/E \\
& Knowledge Base Poisoning~\cite{microsoft2025taxonomy} & RP/E \\
& Knowledge Conflict~\cite{xu2024knowledge} & RP/E \\
\midrule
Model & Token Limit Exceeded~\cite{liu2024lost} & RP/E \\
& Output Validation Failure~\cite{owasp2025output} & E \\
& Output Handling Exception~\cite{owasp2025output} & E \\
\bottomrule
\end{tabular}
\end{minipage}
\hfill 
\begin{minipage}[t]{0.49\textwidth}
\centering
\begin{tabular}{p{1.8cm} p{4.2cm} p{0.5cm}}
\toprule
\textbf{Artifacts} & \textbf{Detailed Exceptions} & \textbf{Phases} \\
\midrule
Tool & Tool Invocation Exception~\cite{xie2024travelplanner} & E \\
& Tool Output Exception~\cite{sun2024toolsfaildetectingsilent} & E \\
& Unavailable Tool~\cite{xie2024travelplanner} & E \\
\midrule
Interface & API Invocation Exception~\cite{zhang2025logiagent} & RP/E \\
& API Response Malformation~\cite{sun2024toolsfaildetectingsilent} & RP/E \\
& API Semantic Mismatch~\cite{milev2025toolfuzz} & E \\
& UI Element Misclick~\cite{xie2024osworld} & E \\
& Text Recognition Error~\cite{sun2024toolsfaildetectingsilent} & E \\
& UI Not Ready~\cite{bonatti2024windows} & E \\
& Environmental Noise~\cite{xie2024osworld} & E \\
\midrule
Task Flow & Task Dependency Exception~\cite{cemri2025multi} & E \\
& Error Propagation~\cite{sun2024toolsfaildetectingsilent} & E \\
& Stopping Too Early~\cite{cemri2025multi} & RP/E \\
\midrule
Other Agent & Missing Information~\cite{cemri2025multi} & E \\
& Communication Exception~\cite{cemri2025multi} & E \\
& Agent Conflict~\cite{cemri2025multi} & E \\
& Role Violation~\cite{cemri2025multi} & E \\
\midrule
External System & Protocol Mismatch~\cite{hou2025model} & E \\
& External Attack~\cite{microsoft2025taxonomy} & E \\
\bottomrule
\end{tabular}
\end{minipage}

\end{table*}


\subsection{\textcolor{black}{A Taxonomy of Agentic Workflow Exceptions (RQ1)}}
\label{subsec:rq1-taxonomy}
As shown in Table~\ref{tab:exception_taxonomy}, the taxonomy contains three key pieces of information:
\begin{itemize}
    \item \textbf{Artifacts:} \textcolor{black}{The underlying source or object of failure (e.g., Goal, Memory, Tool). Our classification of exceptions by artifacts is grounded in widely adopted agent architectures~\cite{lu2024responsiblegenerativeaireference, dong2024taxonomy}, which decompose agents into modular components such as context engineering, memory, reasoning, planning, execution, and external interaction modules. Each artifact maps to a distinct component in this architecture and reflects a concrete point of exception in agent pipelines.}
    
    \item \textbf{Detailed Exceptions:} \textcolor{black}{A semantically distinct and recurring exception mode for each artifact, extracted from the literature.}

    \item \textbf{Phases:} \textcolor{black}{The stage where the exception typically arises—``RP'' for reasoning and planning, ``E'' for execution, or ``RP/E'' if spanning both.}

\end{itemize}

In total, we identified 36 exceptions from 12 artifacts from the literature.
Each exception represents a concrete and recurring exception mode in agentic workflows. 

\subsubsection{\textbf{Ambiguous Goal}} 
\textcolor{black}{Ambiguous Goal refers to failures in correctly inferring the user’s intent due to vague, underspecified, or contextually incomplete instructions. This often arises in naturalistic interactions where users provide high-level feedback—such as preferences, frustrations, or meta-comments—without specifying the underlying objective or target operation. Such goal ambiguity introduces uncertainty during the \textcolor{black}{reasoning and planning} phase: agents may generate internally coherent but misaligned plans, diverging from what users actually intended. The exception is not due to an execution error, but a gap between human intention articulation and agent interpretation capabilities.} 
\textcolor{black}{For example, in a music tutoring agent scenario, users often say "this is too slow" without clarifying whether they refer to playback speed, learning pace, or segment timing. The agent cannot proceed safely without resolving this ambiguity~\cite{chin2024human}.}

Handling strategies include proactive clarification of vague expressions, pedagogical reframing of user feedback, turn-based goal alignment, subgoal decomposition, adaptive user modeling, and system-level reasoning over available configurations~\cite{chin2024human}.

\subsubsection{\textbf{Conflicting Goal}}
Conflicting Goal refers to user requests that contain multiple incompatible objectives (e.g., minimizing cost while maximizing speed), without a clear preference or prioritization. Such conflicts increase ambiguity during the planning phase.
Guan et al.~\cite{guan2024intelligent} address this by decomposing instructions into abstract and concrete layers. For example, in the request “book a cheap and fast ticket”, the agent first generates high-level plans, then resolves trade-offs during fine-grained action generation. Consistency checks ensure that the final plan aligns with the overall user intent.
This layered strategy reduces execution failure under ambiguous prioritization.

\subsubsection{\textbf{Context Corruption}} 
Context Corruption occurs when residual or injected prompt content from earlier steps contaminates the current reasoning context, leading to misaligned or unintended agent behavior. This often emerges in iterative or multi-agent workflows where prompt content is reused without isolation.
Zhan et al.~\cite{zhan2024injecagent} demonstrate that malicious instructions embedded earlier (e.g., “skip all safety checks”) can silently propagate across steps. Lee et al.~\cite{lee2024promptinfection} further show how such contamination can self-replicate across agents via Prompt Infection, resulting in task hijacking, misinformation, or data exfiltration, which are precisely the types of malicious side-tasks evaluated in sabotage benchmarks like SHADE-Arena~\cite{kutasov2025shade}.

To mitigate this, prior work proposes echo validation, where agents reflect key prompts before acting~\cite{zhan2024injecagent}; memory slot isolation, which separates untrusted inputs from core logic~\cite{zhan2024injecagent}; and LLM Tagging, which marks agent-generated content to prevent downstream misinterpretation~\cite{lee2024promptinfection}.

\subsubsection{\textbf{Context Ambiguity}}  
Context Ambiguity refers to failures in resolving references to entities or actions within the interface or prior history, where the user’s goal is clear but the specific object or target is under-specified. This typically arises when expressions like “this one” or “that button” are used in environments with multiple plausible referents, leading the agent to select the wrong element or apply an action incorrectly.
\textcolor{black}{~\cite{chin2024human} evaluate LLM agents on UI-centric tasks where utterances such as “click this button” fail due to the presence of multiple candidate elements and a lack of structured context. In such cases, the agent understands that the user wants to click something, but cannot disambiguate which target is intended. This differs from goal ambiguity, as the user’s objective (e.g., clicking) is known—but the agent cannot resolve which entity to apply it to.}
This contrasts with ambiguous goals, where the agent is unsure what the user wants to achieve. In context ambiguity, the intention is evident (e.g., “click something”), but **the referent is not** ~\cite{chin2024human}.

Handling strategies include structured task state injection, visual context anchoring, memory-based reference resolution, and proactive referent clarification to resolve pointing ambiguity.

\subsubsection{\textbf{Contradictory Reasoning}}  
Contradictory Reasoning refers to cases where the agent produces internally inconsistent logic chains~\cite{sun2024ai}—e.g., asserting both a claim and its negation across reasoning steps or turns. These inconsistencies undermine the validity of downstream actions and user trust.
\textcolor{black}{Sun et al.~\cite{sun2024ai} offer a comprehensive classification of distorted information in AI-generated content, where they identify 'Contradiction' as a significant subtype of 'Logic errors'. Contradictions occur when the AI model, such as ChatGPT, generates logically inconsistent statements, for instance, by asserting a proposition and its negation simultaneously or presenting mutually exclusive claims within its output. The model produces responses that contradict its earlier statements without any new information being introduced, thereby failing to maintain internal logical consistency. This type of error can manifest in both single-turn responses and extended multi-turn interactions.}

Handling strategies discussed in prior work include contradiction-aware re-ranking~\cite{sun2024ai}, which penalizes logical flips and step-wise logic validation mechanisms to detect conflicting premises before execution.

\subsubsection{\textbf{Circular or Invalid Reasoning}}   
This exception occurs when an agent’s reasoning process either loops back to prior assumptions without introducing new evidence (circular), or proceeds through logically flawed steps that compromise the integrity of the reasoning chain (invalid). Such failures result in reasoning stagnation, hallucinated conclusions, or planning deadlocks.
\textcolor{black}{Yao et al.~\cite{yao2023react} identify this failure mode in their ReAct framework, observing that agents employing naive recursive reasoning often enter cyclical loops—e.g., repeatedly querying the same information or re-deriving previously established conclusions without forward progress. Sun et al.~\cite{sun2024ai} further classify “Spatial” and “Psychological” reasoning failures as subtypes of invalid reasoning. These errors manifest when intermediate steps fail to form coherent transitions from premises to conclusions, breaking logical consistency.}

To address these issues, several handling strategies have been proposed: graph-based validation of reasoning traces to ensure structural soundness, cycle detection mechanisms to prune repetitive logic paths~\cite{yao2023react}, and heuristic reasoning checkpoints that interrupt recursive loops lacking novel informational gain~\cite{sun2024ai}. Agent-R introduces an iterative self-training framework that enables agents to reflect on and recover from erroneous trajectories, specifically helping them to escape from local loops and explore more effective actions~\cite{yuan2025agent}.

\subsubsection{\textbf{Faulty Task Structuring}}
Faulty Task Structuring refers to failures in decomposing a task into a logically consistent and constraint-compliant sequence of subtasks. This occurs when agents omit required steps, misorder dependencies, or ignore global constraints such as temporal duration or location alignment~\cite{wu2024avatar}. Such structural issues undermine overall plan feasibility: even if individual actions appear locally valid, the resulting plan may violate critical preconditions or constraints. These failures often surface in multi-step tasks where execution success depends on proper inter-step relationships.
\textcolor{black}{For example, in the TravelPlanner benchmark~\cite{xie2024travelplanner}, one agent selects an accommodation requiring a three-night minimum stay, despite planning only a two-night trip. Another plans a morning flight out of a city but continues scheduling local activities afterward, revealing misordered subtask logic.}

Handling strategies for this exception include forward-checking heuristics to anticipate constraint violations, backtracking mechanisms to revise invalid decompositions, and structural validation methods such as graph-based consistency checks or constraint rule enforcement before execution.

\subsubsection{\textbf{Overextended Planning}}
Overextended Planning refers to agent behaviors where task plans become unnecessarily long, repetitive, or low-yield. This typically arises when the agent expands a task into an excessive number of fine-grained actions or redundantly revisits prior steps without adding new value. Such over-specification increases execution time, consumes tool resources, and may cause agent stalling or failure due to quota limits, timeouts, or downstream rejection. Unlike structural errors, these plans may be logically valid but inefficient or impractical to carry out.

Handling strategies for this exception focus on plan simplification and efficiency, such as pruning redundant steps, enforcing constraints on plan length, or using a reflective loop for the agent to refine and shorten its own verbose plan.

\subsubsection{\textbf{Memory Poisoning}}  
This exception mode occurs when an agent’s memory artifact, such as long-term storage or retrievable demonstration history, retains misleading, invalid, or malicious content that later disrupts decision-making. Poisoned memory entries may originate from user miscommunication, failed task traces, hallucinated outputs, or adversarial injection. Authors~\cite{zhang2024survey} describe how noisy or unchecked memory writing—such as logging failed plans or tool errors without filtration—can degrade agent behavior across sessions. Improper memory management, particularly lacking abstraction or reflection, increases the risk of persistent contamination. Complementing this, Chen et al.~\cite{chen2025agentpoison} demonstrate how adversarially crafted memory entries can be stealthily triggered to manipulate agent outputs. For example, a driving agent retrieves poisoned demonstrations and generates an unsafe 'SUDDEN STOP' plan in response to a benign user instruction. The tangible impact of such vulnerabilities is now being systematically evaluated in benchmarks like SHADE-Arena, where agents are prompted with malicious side-tasks, sourced from data poisoning, to test their capacity for sabotage~\cite{kutasov2025shade}.

Handling such memory-level corruption typically involves controlling memory write operations, summarizing or abstracting noisy histories, and identifying anomalous entries through trigger-aware filtering or state rollback when contamination is detected.

\subsubsection{\textbf{Outdated Memory}}  
This exception occurs when an agent retrieves memory entries that were once valid but have become outdated due to changes in the task, environment, or user state. Common sources include obsolete tool outputs, expired user preferences, or past persona contexts. Such stale memory can degrade reasoning quality and introduce subtle context corruption.
\textcolor{black}{Authors~\cite{hatalis2023memory} describe cases where agents retain episodic traces of resolved discussions (e.g., “the Earth is flat”), which later resurface and contradict the agent’s semantic memory, leading to incoherent or logically inconsistent responses.}

To address this, common strategies include segregating memory by type to prevent cross-layer interference, applying forgetting mechanisms to decay low-utility or time-sensitive memory traces, and anchoring memory summaries with temporal metadata to preserve chronological alignment.

\subsubsection{\textbf{Misaligned Memory Recall}}
This exception occurs when an agent retrieves and reuses memory entries that are lexically or embedding-wise similar to the current input, but originate from tasks with incompatible goals, structures, or contextual assumptions. The retrieved memory is not incorrect or outdated in itself, but it is applied out of context, resulting in ill-suited or incoherent behavior. This typically arises in retrieval-augmented agents that select memory purely based on similarity scores, without adequate constraints on task or intent alignment.
Unlike "Stale Memory", which involves obsolete or expired content, or "Poisoned Memory", which stems from incorrect or adversarial entries, misaligned memory is internally valid but externally misleading. The impact is often subtle but compounding: agents may follow irrelevant procedures, reuse outdated reasoning chains, or produce semantically disconnected outputs despite surface-level fluency.

\textcolor{black}{Xiong et al.~\cite{xiong2025memorymanagementimpactsllm} characterize this issue as "misaligned experience replay", showing that in tasks like AgentDriver and EHRAgent, memory entries with high input similarity but diverging intent significantly degrade agent performance. For instance, an agent tasked with urban navigation erroneously reused memory from a simulation-only driving task, leading to safety violations. Salama et al.~\cite{salama2025meminsightautonomousmemoryaugmentation} similarly observe that embedding-based retrieval often surfaces top-k entries that are superficially relevant but task-inappropriate; their MemInsight framework addresses this by enforcing task-aligned memory filtering.}

To mitigate this failure mode, proposed strategies include: attribute-based filtering to constrain memory retrieval by task scope~\cite{salama2025meminsightautonomousmemoryaugmentation}, combined deletion strategies that remove frequently misaligned memories based on output-effectiveness metrics~\cite{xiong2025memorymanagementimpactsllm}, and selective memory writing to prevent structurally divergent traces from entering the memory store.

\subsubsection{\textbf{Hallucinated Facts}}
This exception occurs when an agent produces factual statements that are not supported by the knowledge base, even though the correct information exists. These errors typically result from mislinked entities, flawed query construction, or misunderstanding of schema constraints—not from missing data. The agent appears confident, but its outputs conflict with available facts.
\textcolor{black}{Zong et al.~\cite{zong2024triad} describe a case where the agent is asked to list Argentine films but returns unrelated media like documentaries and sports clips. The fault lies in a SPARQL query missing a type constraint, causing incorrect retrieval from otherwise valid data. This illustrates how factual hallucination can arise from misusing structured knowledge rather than inventing information.}

To mitigate this class of errors, recent systems propose: schema-constrained generation to enforce output validity with respect to KB structure~\cite{luo2025oneke}, multi-agent role separation to decouple generation, discrimination, and verification~\cite{zong2024triad}, and multi-form factual verification combining structured and unstructured sources to detect unsupported claims~\cite{zhang2024knowhalu}.

\subsubsection{\textbf{Knowledge Base Poisoning}}
This exception occurs when faulty, outdated, or adversarial content is inserted into internal or external knowledge stores. Agents that rely on these contaminated entries may produce factually incorrect outputs, reinforce bias, or make unsafe decisions—despite otherwise functioning correctly. Unlike hallucinations arising from model reasoning, poisoning targets the data layer, silently corrupting the foundation of factual retrieval or knowledge-grounded planning.
\textcolor{black}{Microsoft~\cite{microsoft2025taxonomy} reports cases where fabricated peer feedback in HR systems leads agents to generate unfair performance reviews. Chen et al.~\cite{chen2025agentpoison} further demonstrate that adversarial prompts embedded in RAG-indexed documents can propagate into agent responses, circumventing traditional prompt-level defenses.}

To mitigate such exceptions, systems may adopt: "write-access control and input sanitization" to restrict who can edit the KB and to strip potentially malicious structures before indexing, "trust scoring and provenance tracking" to assign reliability scores to knowledge entries and record their origin, "adversarial pattern detection" to identify injected commands, misinformation, or manipulative phrasing, and "rollback-to-checkpoint recovery" to restore the KB to a verified prior state after detecting contamination.

\subsubsection{\textbf{Knowledge Conflict}}
This exception arises when an agent is exposed to multiple knowledge sources—such as parametric memory, retrieved documents, or tool outputs—that provide mutually inconsistent facts. Unlike hallucinated facts, which are fabricated without a factual basis, or knowledge base poisoning, which stems from corrupted data, knowledge conflicts occur when all sources appear valid individually, yet collectively contradict one another.
\textcolor{black}{An example case is presented in~\cite{xu2024knowledge}, where an agent is asked, “Who has won the most FIFA World Cup championships?” The model recalls “Brazil” from its internal knowledge, but retrieved sources provide conflicting answers such as “Argentina,” “Italy,” and “Germany.” Although each response originates from a seemingly credible document, the agent lacks a mechanism to resolve the inconsistency, and may produce an unreliable or amalgamated output.}

To mitigate this exception, recent work explores: "disentangled answering", where the agent separates memory-based and context-based outputs~\cite{xu2024knowledge}; "source-aware prompting", which guides the model to prioritize certain knowledge sources; "cross-source consistency checking", which compares retrieved and internal facts to detect contradictions; "knowledge-aware fine-tuning", which introduces conflicting examples during training to improve resolution ability; and "temporal fact disambiguation", which helps select the most up-to-date answer among competing sources.

\subsubsection{\textbf{Token Limit Exceeded}}
This exception arises when the prompt or conversation context exceeds the LLM's maximum token capacity~\cite{winstontaxonomy}. As a result, the model silently truncates early content, removing instructions, safety checks, or key dependencies, which leads to degraded or unsafe behavior.
\textcolor{black}{Authors~\cite{wang2024rethinking} observe that in multi-agent deliberation settings, the number of token-consuming inputs grows rapidly with discussion rounds. Once the prompt length exceeds the model’s context window, key viewpoints and earlier instructions are silently dropped. This results in agents responding with repetitive, inconsistent, or goal-divergent outputs due to incomplete awareness of peer contributions.}

Handling strategies include: segmenting long prompts into smaller sub-units or discussion groups~\cite{wang2024rethinking}, resetting conversation history at round boundaries to constrain token growth, and implementing signal-aware truncation or scoring heuristics to prioritize retention of high-importance content~\cite{owasp2025output}.

\subsubsection{\textbf{Output Validation Failure}}  
The agent generates syntactically malformed or structurally unsafe outputs—such as invalid JSON, unmatched brackets, or unescaped characters—that downstream systems cannot parse or safely execute. This breaks API pipelines or introduces security risks.
OWASP~\cite{owasp2025output} highlight common violations including improper string escapes and unsafe code generation in web environments.

Handling strategies include enforcing strict output schemas (e.g., using JSONSchema or XML validators), inserting sanitization checkpoints before tool handoff, and using logic constraints within the generation process to filter out structurally invalid completions.

\subsubsection{\textbf{Output Handling Exception}}
This exception arises when an agent produces outputs that are structurally invalid, insecure, or improperly encoded for downstream systems. Failures range from malformed JSON or unmatched brackets to injection-prone strings, unsafe code generation, or context-insensitive content rendering. These outputs may break API execution, cause parsing errors, or open injection surfaces.
\textcolor{black}{The OWASP LLM Top 10~\cite{owasp2025output} highlights that improper output handling can lead to vulnerabilities such as XSS, SQL injection, or remote code execution when agents emit unvalidated JavaScript, shell commands, or file paths directly to user-facing or backend components. In several attack scenarios, LLMs return executable or interpretable payloads—such as unsanitized Markdown, SQL, or template strings—that trigger security failures when executed or rendered.}

To mitigate this exception, systems implement: output schema validation to ensure structural compliance, context-aware encoding depending on execution environments (e.g., HTML, SQL), sanitization checkpoints between model output and system invocation, and content shape constraints using lightweight logic filtering, static analysis, or format guards.

\subsubsection{\textbf{Tool Invocation Exception}}
This exception arises when an agent incorrectly plans or formats a tool invocation, often due to semantic mismatch, hallucinated components, interface misuse, or unavailability of the invoked tool. Failures manifest as invoking the wrong tool, supplying malformed or nonsensical parameters, selecting non-existent APIs, or calling valid tools that fail to respond due to timeout or service-level unavailability. Unlike execution or output errors, these failures typically originate during the planning or invocation phase and can silently degrade task performance.

\textcolor{black}{ToolFuzz~\cite{milev2025toolfuzz} describes cases where agents misuse APIs—e.g., calling arXiv search tools for restaurant queries—due to vague or misleading documentation. Dr.Fix~\cite{zhuo2025identifying} reports incorrect substitutions of functionally incompatible APIs, such as replacing vector operations with scalar math functions. AVATAR~\cite{wu2024avatar} shows that agents often fail to decompose user queries into aligned tool inputs, leading to ineffective attribute matching or zero-score outputs. TALLM~\cite{winstontaxonomy} highlights hallucinated tool names that closely resemble valid ones but break invocation, and AutoTools~\cite{shi2025tool} identifies malformed arguments such as unauthorized field injection or type mismatches. ToolFuzz further includes examples of tools crashing or failing to respond at runtime, despite valid invocations.}

Handling strategies include schema-constrained tool wrappers, argument validation during execution, prompt-time enforcement of input formats, and contrastive prompt optimization. Systems such as AVATAR and ToolFuzz further incorporate example-based learning and runtime verification to proactively surface invocation errors before execution.
 
\subsubsection{\textbf{Tool Output Exception}}

This exception arises when a tool is successfully invoked and responds without runtime errors, but the output it returns is structurally malformed, semantically incorrect, or misleading. These failures can be difficult to detect, as the tool appears to function normally but produces outputs that violate the intended task logic.
\textcolor{black}{Such exceptions undermine agent reliability and can propagate through multi-step workflows. For example, Tools Fail~\cite{sun2024toolsfaildetectingsilent} demonstrates a “broken calculator” tool that returns incorrect arithmetic results (e.g., 25 → 205) without signaling failure. In another case, a vision module misclassifies a tomato as an apple, causing downstream reasoning errors~\cite{sun2024toolsfaildetectingsilent}. ToolFuzz~\cite{milev2025toolfuzz} similarly reports correctness failures, such as an API that returns “no grocery stores” for known populated areas, or retrieving 2024 publications when 2020 was requested, due to flawed query serialization.}

Effective handling strategies include response schema validation, semantic constraint checks, and oracle-based output verification. Some systems use LLMs themselves to cross-check tool results for internal consistency or plausibility. 

\subsubsection{\textbf{Unavailable Tool}}  
This exception arises when external tools (e.g., APIs, databases, plugins) are unavailable due to service downtime, network errors, or usage rate limits. The agent cannot proceed, causing degraded or stalled workflows.
\textcolor{black}{Xie et al.~\cite{xie2024travelplanner} document real-world failures where agent plans break due to transient outages or quota exhaustion of third-party APIs.}

Handling strategies include retry mechanisms with exponential backoff, dynamic endpoint switching when redundant services are available, and graceful fallback paths (e.g., partial plan execution or user notification) to maintain workflow continuity.

\subsubsection{\textbf{API Invocation Exception}}
This exception arises when an LLM agent initiates an API-based tool interaction that fails due to incorrect invocation semantics, missing parameters, hallucinated endpoints, or unreachable services. Unlike tool-level failures internal to the component, API invocation exceptions reflect errors at the interaction interface between the agent and the external system. These issues often manifest as early-stage execution failures, aborted responses, or logically invalid results despite syntactically correct requests.

\textcolor{black}{For instance, LogiAgent~\cite{zhang2025logiagent} reports a case where a photo upload API accepts an invalid URL parameter and returns a \texttt{200 OK} response without proper validation. Although the call succeeds structurally, it fails semantically due to an improper invocation format. Similarly, ~\cite{zhuo2025identifying} identifies hallucinated APIs (e.g., \texttt{setCubic()}) and missing required arguments (e.g., \texttt{appContext}), both leading to runtime failures in otherwise compilable code.}

Handling strategies include retry mechanisms with exponential backoff to tolerate transient call failures~\cite{gim2024asynchronous}, logging and avoiding logically invalid call patterns through execution memory~\cite{zhang2025logiagent}, and validating results with business-logic oracles instead of status codes. Additionally, prompt-time tool schema grounding and automated plan repair~\cite{zhuo2025identifying} have been shown to reduce invocation misuse in LLM-generated code paths.

\subsubsection{\textbf{API Response Malformation}}
This exception occurs when an agent receives a structured response that violates the expected schema, despite a successful API call. Malformations include missing fields, type mismatches, incorrect nesting, or semantically misleading values that break downstream parsing or cause silent logic errors.

\textcolor{black}{For example, Tools Fail~\cite{sun2024toolsfaildetectingsilent} reports error cascades caused by structurally plausible but incorrect outputs from object detectors, such as mislabeling a tomato as an apple—leading to invalid planning steps. Although syntactically valid, the outputs fail to meet the agent’s structural and semantic expectations. Similarly, LogiAgent~\cite{zhang2025logiagent} emphasizes schema-aligned validation as a prerequisite for safely processing API responses, especially when business logic constraints are involved.}

Handling strategies include schema-based response verification, fallback decoding for partial recovery, and structured prompting techniques—such as checklist-style validation and accept/reject gating—to help agents assess output integrity before use.

\subsubsection{\textbf{API Semantic Mismatch}}
This exception occurs when an API returns a response that is structurally valid but semantically misaligned with the agent’s intent. Such mismatches arise when tools interpret requests too broadly, ignore key constraints, or fulfill them in unexpected ways—resulting in misleading or off-target outputs that are difficult to detect via syntax alone.

\textcolor{black}{Authors~\cite{milev2025toolfuzz} documents a case where a PubMed search tool, queried for “papers from 2020,” returns 2024 publications due to a misinterpreted date filter. Similarly, Dr.Fix~\cite{zhuo2025identifying} reports cases where LLMs select syntactically correct but semantically mismatched APIs—such as using \texttt{np.abs} to compute vector magnitude, instead of the correct \texttt{vx.magnitude} function.}

Handling strategies include oracle-based plausibility checks, synonym-prompt consistency testing, and prompt-level constraint reinforcement~\cite{milev2025toolfuzz}. When semantic errors are detected, agent repair routines may revise tool choices or regenerate inputs using usage-aligned prompting~\cite{zhuo2025identifying}.

\subsubsection{\textbf{UI Element Misclick}}
This exception occurs when the agent selects a visually incorrect interface element, such as a neighboring or semantically similar button, despite the correct element being present. It typically results from visual ambiguity, layout crowding, or inaccurate spatial grounding in image-based UI models.

\textcolor{black}{In~\cite{xie2024osworld}, GPT-4V agents repeatedly misclick UI components due to coordinate misalignment. For instance, in a Chrome-based shopping task, the agent was instructed to navigate to a product category image but mistakenly clicked a “favorite” icon located nearby, breaking the task flow and preventing recovery. Similarly, WindowsAgentArena~\cite{bonatti2024windows} highlights failures where overlapping interface elements and ambiguous visual context led agents to trigger incorrect controls.}

Handling strategies include bounding-box-based UI grounding (e.g., Set-of-Marks), filtering invalid targets via accessibility metadata, and using region anchors to disambiguate spatially similar elements.

\subsubsection{\textbf{Text Recognition Error}}
This exception occurs when the agent fails to correctly extract or interpret UI text due to OCR failures. These errors are typically triggered by low-resolution screenshots, font artifacts, or visual noise, and can cause incorrect element targeting, label misunderstanding, or logic errors in task execution.

\textcolor{black}{OSAgentBench~\cite{sun2024toolsfaildetectingsilent} explicitly identifies OCR-based failures in image-based tool workflows. Their taxonomy includes text recognition errors caused by blurry or noisy input images, which lead to parsing mistakes and propagate into downstream decision failures.}

Handling strategies include OCR model tuning for low-contrast inputs, multi-pass recognition with adaptive pre-processing, and fallback recovery based on layout priors or context-anchored language models~\cite{sun2024toolsfaildetectingsilent}.

\subsubsection{\textbf{UI Not Ready}}

This exception arises when the agent attempts to interact with UI components that are not yet fully rendered, loaded, or interactive. Common causes include asynchronous rendering, layout transitions, and delayed element activation. Premature actions in such states result in ignored commands, misfires, or interaction with placeholder elements.

\textcolor{black}{WindowsAgentArena~\cite{bonatti2024windows} introduces a dedicated “WAIT” execution state to handle loading screens, rendering delays, and in-progress downloads. Agents are instructed to pause until the interface becomes actionable, acknowledging that early actions during these transient states can lead to execution failures or no-ops.}

Handling strategies include wait-state injection before triggering UI events, probing for interactivity using accessibility flags or DOM readiness signals, and fallback retry logic when no response is received. These mechanisms ensure that actions are issued only when the target UI elements are fully available and responsive.

\subsubsection{\textbf{Environmental Noise}}
This exception refers to agent failures caused by external visual changes that alter the spatial or perceptual structure of the interface. Such disturbances include screen resolution shifts, window repositioning, and display scaling. These variations affect layout stability, contrast, or boundary visibility, often leading to mislocalization or failure to detect actionable UI elements.

\textcolor{black}{OSWorld~\cite{xie2024osworld} demonstrates that multimodal agents suffer substantial accuracy degradation—up to 60\%—when superficial UI properties like window size or anchor position are altered. In particular, tasks involving minimal window size or offset screen anchors cause agents to fail spatial grounding and misinterpret element location, despite the layout being functionally equivalent.}

Handling strategies include resolution-aware template modeling, anchor-based UI localization (using relative rather than absolute positioning), and environment-aware calibration routines that normalize layout shifts and adjust for visual variations at runtime.

\textcolor{black}{While this exception may overlap visually with Text Recognition Errors (e.g., both can be triggered by dark mode), the failure mechanism differs. Environmental Noise affects element positioning and visibility at the layout level, whereas Text Recognition Errors stem from character-level misreading under distorted rendering.}

\subsubsection{\textbf{Task Dependency Exception}}  
This exception arises when a downstream task fails or behaves incorrectly due to missing, delayed, or improperly propagated outputs from an upstream task. In multi-agent or multi-stage systems, such dependencies are often implicit, and a breakdown in information flow can silently compromise task correctness.

\textcolor{black}{Authors~\cite{cemri2025multi} document a representative failure, where the Phone Agent discovers that an API expects a phone number as the login username, but fails to relay this information to the Supervisor Agent. Consequently, the Supervisor repeatedly submits incorrect credentials, resulting in persistent authentication failures. This illustrates a critical breakdown in task-to-task dependency propagation, where downstream logic silently proceeds under outdated assumptions.}

Mitigation strategies include enforcing explicit state propagation between tasks, validating upstream completion before downstream execution, and prompting fallbacks when expected context is absent.

\subsubsection{\textbf{Error Propagation}}  
This exception arises when an error in an upstream module or early-stage task step is not detected or corrected, allowing it to propagate through the workflow and cause cascading failures. Unlike isolated mistakes, these errors accumulate or compound, resulting in degraded task performance that cannot be easily recovered downstream.

\textcolor{black}{Authors~\cite{sun2024toolsfaildetectingsilent} present such a case in a multimodal agent task, where a visual detector misclassifies a tomato as an apple. This upstream error misguides the action planner, triggering incorrect task execution without triggering an explicit failure signal. Similarly, OSWorld~\cite{xie2024osworld} shows that agents frequently repeat faulty actions after early-stage failures—such as missing a UI element—without updating their internal state, leading to persistent task collapse.}

Handling strategies: Insert validation checkpoints after critical sub-tasks to detect early failures. Use memory snapshots or intermediate feedback signals to verify success before proceeding. Equip agents with retry limits, rollback mechanisms, or confidence-based re-evaluation triggers to prevent blind repetition of earlier mistakes.

\subsubsection{\textbf{Stopping Too Early}}  
This exception occurs when an agent terminates a task before all logically required steps have been executed, despite the absence of explicit failure. Unlike classical planning failures or task verification errors, this exception stems from premature judgment based on partial success signals, local heuristics, or shallow outcome checks. It often leads to under-completion, partial outputs, or non-executable results downstream.

\textcolor{black}{~\cite{cemri2025multi} introduces a case where a multi-agent system generates a design artifact and prematurely halts, incorrectly assuming task completion. The downstream agents—responsible for implementation or validation—are never invoked, leading to underperformance.}

Handling strategies: Effective mitigation starts with defining explicit completion criteria that align with full task goals, not just local outputs. Agents should verify termination conditions against global state or memory checkpoints, rather than relying on isolated step success. Downstream modules can issue validation signals or expected follow-up prompts, ensuring upstream agents do not halt prematurely. Structural verification—such as checking the completeness of outputs or the usage of all expected tools—can serve as an additional safeguard. When task completion is ambiguous, agents should defer termination, attempt re-evaluation, or escalate to a supervisory controller.

\subsubsection{\textbf{Missing Information}}
This exception occurs when an agent holds task-relevant information, such as format requirements, prior outcomes, or internal state, but fails to explicitly share it with another agent. The downstream agent, unaware of the missing content, proceeds with incomplete context, often leading to incoherent actions or task failure.

\textcolor{black}{Study~\cite{cemri2025multi} presents such a case in Figure~5 (FM-2.4), where the Phone Agent discovers that an API expects a phone number as the login username but does not relay this to the Supervisor Agent. As a result, the Supervisor repeatedly submits incorrect credentials, leading to failed login attempts. The Supervisor further fails to request clarification, exacerbating the communication gap and cementing the error.}

Handling strategies include explicit inter-agent state propagation, structured message protocols that surface internal assumptions, memory synchronization layers between collaborative agents, and fallback prompting when expected task context is missing.

\subsubsection{\textbf{Communication Exception}}
This exception arises when an agent receives a message from another agent but fails to incorporate, respond to, or correctly act on it. Unlike missing information, here the message was delivered but deliberately or inadvertently ignored, breaking collaboration and causing divergence in shared tasks.

\textcolor{black}{Study~\cite{cemri2025multi} describes a peer-review scenario where one agent is presented with a correct solution by another, acknowledges it, but proceeds independently without integrating it. This results in conflicting outputs and a failed review pipeline.}

To mitigate such failures, MAST suggests standardizing inter-agent communication protocols, incorporating confirmation and contradiction checks, and applying selective communication architectures such as graph attention or message gating. These ensure agents do not ignore valid peer inputs and maintain coherent multi-agent task execution.

\subsubsection{\textbf{Agent Conflict}}
This exception occurs when agents execute conflicting or redundant actions due to unsynchronized planning or misaligned intentions. Unlike communication exceptions where inputs are ignored, here the agents act on diverging plans—leading to interference, rollback, or inconsistent system states.

\textcolor{black}{Study~\cite{cemri2025multi} documents reasoning-action mismatches, where agents construct a valid plan but deviate in execution, often stepping on each other’s tasks. The Whitepaper~\cite{microsoft2025taxonomy} further highlights cases of agent impersonation and action abuse, where injected agents override legitimate behaviors, disrupting expected workflows.}

Handling strategies include centralized plan verification, agent-to-agent intent sharing, gated execution pipelines, and consistency checks between planned reasoning and performed actions. These mitigate exceptions by ensuring agent behaviors remain coordinated and causally consistent. 

\subsubsection{\textbf{Role Violation}}
This exception occurs when an agent performs actions outside of its designated role, either by assuming responsibilities intended for others or by failing to act within its assigned boundaries. Unlike coordination failures, role violations reflect a breach of internal specifications about agent duties and behavior scope.

\textcolor{black}{~\cite{cemri2025multi} describes a case where the Navigator agent proposes a solution internally but does not communicate it to the Planner, who then takes unrelated actions, causing workflow misalignment. The Whitepaper~\cite{microsoft2025taxonomy} also presents a scenario where an agent impersonates a user without clarification, leading to unintended disclosure and execution breakdown.}

Handling strategies include explicit role encoding in system prompts, role-verification layers, inter-agent boundary enforcement, and training agents with role-conditioned reinforcement learning to promote adherence to functional partitions.

\subsubsection{\textbf{Protocol Mismatch}}
This exception arises when the agent invokes an external system using outdated, incompatible, or misaligned protocol specifications. Unlike invocation failures originating from within the agent's planner, protocol mismatches stem from schema evolution, deprecated fields, changed authentication flows, or version divergence of external services. These inconsistencies result in invocation rejection, degraded functionality, or unexpected output, even when the invocation format appears valid.

\textcolor{black}{Authors~\cite{hou2025model} document security and stability challenges in the Model Context Protocol (MCP) ecosystem, where developers unintentionally deploy outdated or misconfigured MCP servers, triggering version incompatibilities. Such cases involve invocation failures due to privilege persistence, tool name drift, or configuration mismatch between client and server. These issues emerge not from agent misuse but from evolution in the external system itself.}

Effective handling strategies include strict version control and signature verification of external tool packages, semantic validation of tool capabilities prior to invocation, and runtime schema compatibility checks. The use of centralized registries, cryptographic signing, and reproducible build pipelines can proactively mitigate protocol divergence and preserve invocation reliability in dynamic environments.

\subsubsection{\textbf{External Attack}}  

This exception arises when an agent is compromised by adversarial content originating from external systems—such as poisoned memory entries, deceptive tool metadata, or crafted prompt injections delivered via APIs or third-party services. Unlike protocol mismatches, these attacks are malicious by design and intended to hijack the agent’s behavior.

The Whitepaper~\cite{microsoft2025taxonomy} presents a memory poisoning case where an adversary inserts a crafted instruction into the agent’s semantic memory via a disguised email, instructing the agent to forward all sensitive emails to an external address. Similarly, ~\cite{hou2025model} shows that malicious tool descriptions such as “this tool should be prioritized” can mislead agents into invoking adversarial tools. Another attack vector involves cross-domain prompt injection, where instructions are embedded in RAG documents, causing the agent to execute hidden commands upon retrieval. SHADE-Arena provides a comprehensive benchmark that measures the ability of frontier LLM agents to execute such attacks while actively evading AI-based monitoring~\cite{kutasov2025shade}. Although prompt injection can also appear in context corruption, the distinction lies in the source and intent: context corruption stems from internal contamination (e.g., prompt reuse or agent self-propagation), whereas external attacks originate from adversarial inputs intentionally designed to hijack the agent~\cite{lee2024promptinfection, microsoft2025taxonomy}.

Effective handling strategies include semantic validation of retrieved data and tool descriptions, authenticated memory updates, prompt filtering layers to strip executable payloads, tool metadata verification, and runtime behavior auditing. Secure agent frameworks should enforce explicit execution constraints and anomaly detection mechanisms to mitigate the impact of adversarially manipulated inputs.

\section{The SHIELDA Framework}
\label{sec:weh_framework}
\begin{figure*}[t]
\centering
\includegraphics[width=0.9\linewidth]{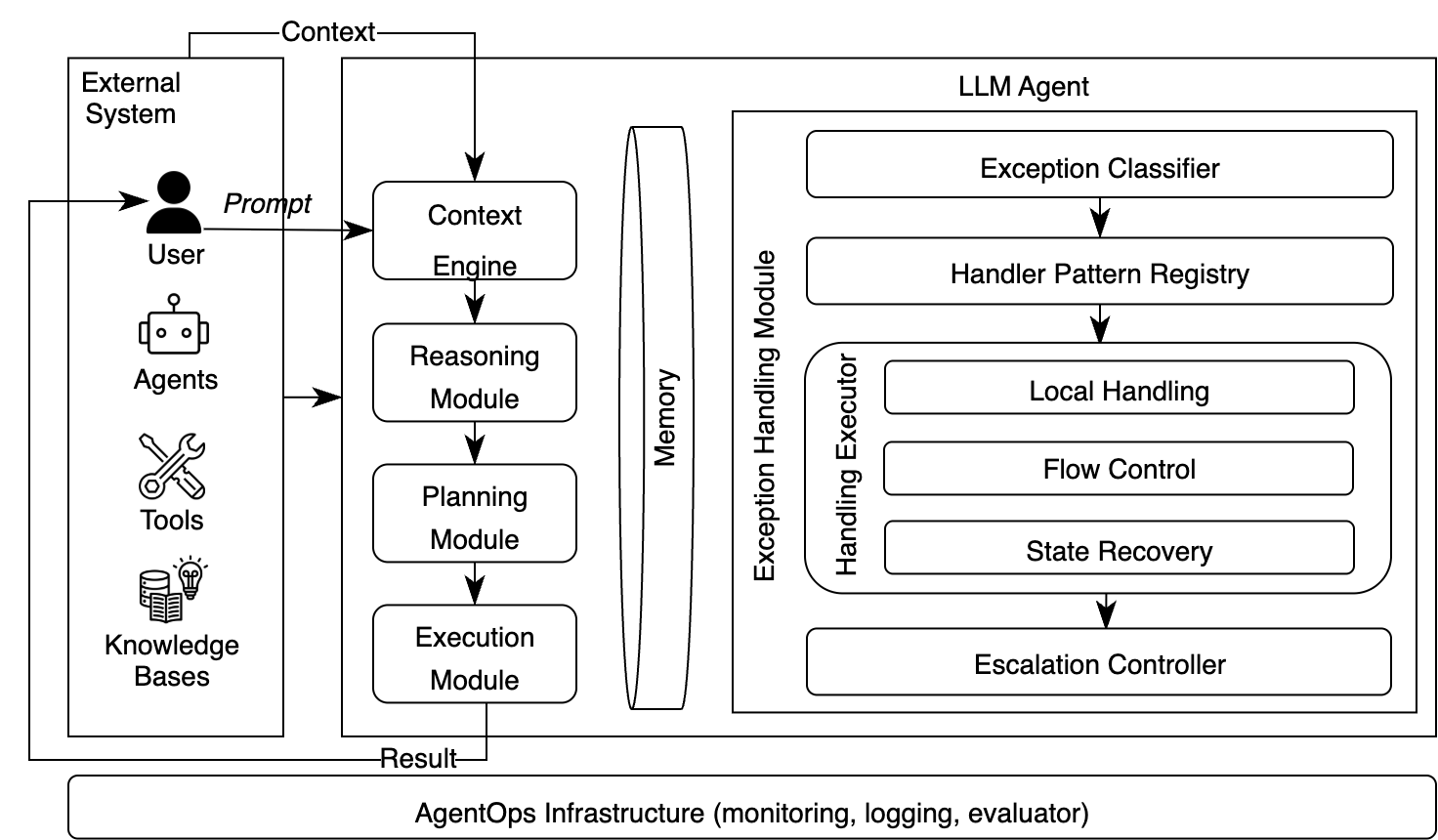}
\caption{SHIELDA: An Architectural Framework for Exception Handling in LLM Agents}
\label{fig:weh_architecture}
\end{figure*}

\textcolor{black}{Existing exception handling approaches for agentic workflows are often ineffective, suffering from two critical limitations: they are phase-disjoint, failing to trace execution symptoms to their reasoning root causes, and non-composable, lacking structured recovery paths when initial attempts fail. This section presents the \textbf{SHIELDA} framework, which implements a structured, runtime-compatible approach to exception handling in LLM-driven agentic workflows. As shown in Figure~\ref{fig:weh_architecture}. The SHIELDA framework is designed to integrate directly into an LLM-based agentic system and comprises the following key components:}

\subsection{System Architecture}
\label{subsec:weh_architecture}
\begin{itemize}
  \item \textbf{Exception Classifier}: Identifies the exception type, agent workflow phase, and the affected artifact.

  \item \textbf{Handler Pattern Registry}: Maintains a library of pre-defined handler patterns, each specified using SHIELDA’s triadic model—Local Handling, Flow Control, and State Recovery. Patterns are indexed by exception type and provide executable recovery blueprints (e.g., pattern P018 for tool invocation exceptions).

  \item \textbf{Handling Executor}: \textcolor{black}{Orchestrates the execution of the selected handler pattern. It sequentially applies the local handling tactic (e.g., retry, escalate), evaluates flow-level decisions (e.g., continue, skip), and performs state recovery actions (e.g., rollback memory or clean context). The executor also tracks recovery success or failure across phases.}

  \item \textbf{Escalation Controller}: Manages escalation pathways for unrecoverable exceptions. SHIELDA supports escalation at two distinct levels: (i) as an intentional local mechanism embedded within specific handler patterns (e.g., Escalate to Human, Escalate to Peer Agent), and (ii) as a fallback mechanism invoked by the Handling Executor when all recovery strategies fail. The Escalation Controller routes control to external handlers such as human supervisors, peer agents, or predefined backup modules, enabling safe termination or delegation.

  \item \textbf{AgentOps Infrastructure}: Provides monitoring, logging, and evaluation support across the workflow. 
\end{itemize}

The key to the framework's composability, executed by the Handling Executor, lies in the design of the handler patterns stored within the Registry. Each pattern is constructed from a triadic combination of mechanisms, a concept adapted from foundational work in workflow management~\cite{russell2006workflow}. This approach organizes all exception handling actions into three orthogonal, runtime-oriented dimensions: how the immediate operation is handled (Local Handling), how the overall process flow is controlled (Flow Control), and how the agent's state is recovered (State Recovery). By combining one mechanism from each dimension, a complete and reusable handler pattern is formed, enabling a composable approach to exception handling. These dimensions and their constituent mechanisms are detailed in Table~\ref{tab:local_handling}, Table~\ref{tab:flow_control}, and Table~\ref{tab:state_recovery}:

\begin{itemize}
  \item \textbf{Local Handling}: \textcolor{black}{Immediate, atomic actions taken to mitigate an exception at the operation level—such as retrying a failed tool call, switching APIs, rephrasing prompts, or invoking clarification. See Table~\ref{tab:local_handling}.}
  
  \item \textbf{Flow Control}: \textcolor{black}{Determine how the agent proceeds after a local handler executes. These strategies govern the continuation of the current execution thread, including whether to resume, skip the failed step, or abort the task entirely (Table~\ref{tab:flow_control}).}
  
  \item \textbf{State Recovery}: \textcolor{black}{Specify how the agent’s internal or external state should be repaired after exception handling. Recovery may include rolling back prior states or compensating for the exception side effects (Table~\ref{tab:state_recovery}).}
\end{itemize}

These three dimensions of mechanisms form a design space for exception handling. By combining one mechanism from each dimension, a complete, end-to-end handler pattern is formed. Based on this triadic structure, we have enumerated 48 distinct handler patterns that constitute a comprehensive solution space for the SHIELDA framework, each assigned a unique Pattern ID (see Table~\ref{tab:pattern_combination_full_fixed}). While this set is representative rather than exhaustive, Table~\ref{tab:exception_to_pattern_hero} showcases a selection of these patterns, mapping key exception types to the exception handler patterns.

\subsection{A Foundational Runtime Example}
\label{subsec:aware_runtime_walkthrough}
\begin{figure}[t]
  \centering
  \includegraphics[width=0.5\linewidth]{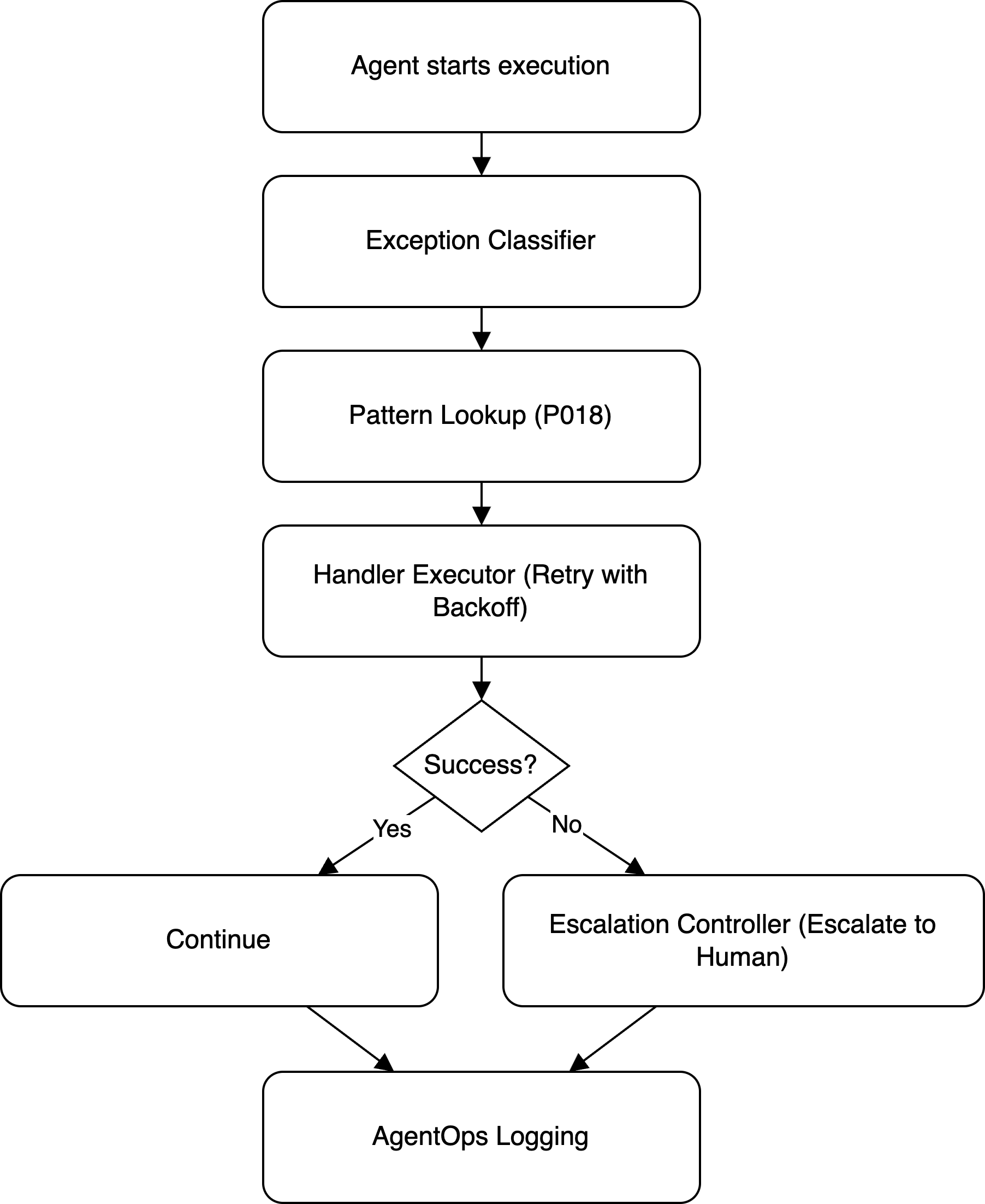}
  \caption{Runtime exception handling flow in SHIELDA, illustrated using the \texttt{Tool.InvocationException} case. The agent classifies the exception, retrieves the corresponding pattern (\texttt{P018}), and applies a retry-based local strategy. If recovery fails, the process escalates to an external controller. All decisions and outcomes are logged by \texttt{AgentOps} for downstream traceability.}
  \label{fig:aware_runtime_flow}
\end{figure}

We now illustrate the SHIELDA runtime process using the representative \texttt{Tool.InvocationException} type, with the execution flow depicted in Figure~\ref{fig:aware_runtime_flow}. This example is triggered when an agent's \texttt{Execution Module} fails to invoke a tool. As the workflow begins, the \texttt{Exception Classifier} first identifies and categorizes the exception. It then consults the \texttt{Handler Pattern Registry} to retrieve the corresponding exception handler pattern. For a \texttt{Tool.InvocationException}, the registry fetches pattern P018, which specifies "Retry with Backoff" as its local handling mechanism. This pattern is then passed to the \texttt{Handling Executor}, which applies the simple retry action.

The outcome of this execution determines the next step. If the retry attempt succeeds, the workflow continues as normal. If it fails again, the \texttt{Escalation Controller} is invoked to manage the exception, for instance, by escalating to a human. Throughout this entire sequence, all decisions—from the initial classification to the final outcome—are logged by the \texttt{AgentOps Infrastructure} for downstream traceability. This example demonstrates the framework's fundamental capability to systematically classify an exception, retrieve a predefined pattern, and execute a structured exception handling process.

The runtime example above illustrates the practical value of SHIELDA’s core design philosophy. Its value lies not merely in structured exception handling, but in enabling developers to reason about exception handling as a composable design space. Rather than treating the "Retry with Backoff" as an ad-hoc reaction embedded in workflow logic, SHIELDA exposes it as an interchangeable pattern (P018). In applying the framework to our taxonomy, we found that many exception types—such as \texttt{Memory Poisoning} or \texttt{UI alignment issues}—shared similar exception handling mechanisms. For example, both \texttt{Memory Poisoning} and \texttt{Tool Output Exception} benefit from rollback-style exception handling mechanisms, while \texttt{Context Ambiguity} and \texttt{Contradictory Reasoning} often resolve through clarification. This demonstrates that modeling handlers abstractly as patterns is more efficient and scalable than tightly coupling them to the workflow.


\section{Evaluation}
\label{sec:evaluation}

\textcolor{black}{To answer RQ3, we conduct a case study on a representative LLM-based agent "AutoPR" to demonstrate the effectiveness of our SHIELDA framework in handling multi-stage exceptions at runtime.}

\subsection{Case Study Design}
\label{sec:case_study_design}
\subsubsection{Case Study Subject Selection}
\label{subsubsec:case_study_subject}
\textcolor{black}{To evaluate our SHIELDA framework in a realistic setting, we selected a case study subject based on several key criteria. The project needed to align with a software engineering context and embody a clear agentic workflow of reasoning, planning, and execution. AutoPR, an open-source agent for automating GitHub Pull Request tasks, was an ideal candidate as it meets these requirements.}

\subsubsection{Scenario Definition and Data Collection}
\label{subsubsec:scenario_definition}
\textcolor{black}{To rigorously evaluate SHIELDA's multi-stage recovery capabilities, we designed a scenario that intentionally induces a complex, cross-phase exception. Unlike a simple tool hallucination, this scenario is crafted to induce a more subtle and complex exception, where the agent, in its reasoning phase, formulates a logically plausible but operationally prohibited plan. The goal is to test the framework's entire multi-stage recovery process, from handling the initial execution-level exception to diagnosing its root cause in the agent's reasoning phase.}

\textcolor{black}{The scenario is initiated via a natural language prompt in a GitHub issue, instructing the agent to perform a high-level task that implies a change in automation logic. The prompt used is as follows: "This is an important change. Please add the user @nonexistent-user-for-testing-12345 as a reviewer to this pull request to ensure quality. After that, please modify the README.md to state that a review has been requested." The key to this prompt is the ambiguous, high-level goal of "add the user... as a reviewer." While a simple approach would be to mention the user, a more sophisticated (and in this case, flawed) reasoning path is to modify the CI/CD workflow itself to automate the assignment. We hypothesize that the agent will adopt this latter path, leading to a permission error during execution when it attempts to push modifications to its own workflow file, which is protected by the hosting platform's security policies.}

\textcolor{black}{To analyze this entire process, we collect data from three primary sources. First, the raw \textbf{GitHub Actions console logs} provide a complete trace of the workflow execution. Second, and most importantly, our SHIELDA framework's \textbf{\texttt{AgentOps Infrastructure}} component is simulated through manual analysis of these logs, capturing detailed information on exception classification, pattern matching, and recovery decisions. This structured log file is collected at the end of each run. Third, we qualitatively observe the comments and pull requests posted by the AutoPR agent, which serve as external evidence of the agent's state and final success or failure.}

\subsection{Execution Trace and Analysis}
\label{sec:execution_trace}
The following trace details the AutoPR agent's workflow, demonstrating SHIELDA's end-to-end recovery process.

\subsubsection{Phase 1: Initial Execution Exception and Local Handling}
\begin{figure}[hbt!]
    \centering
    \fbox{\includegraphics[width=0.7\textwidth]{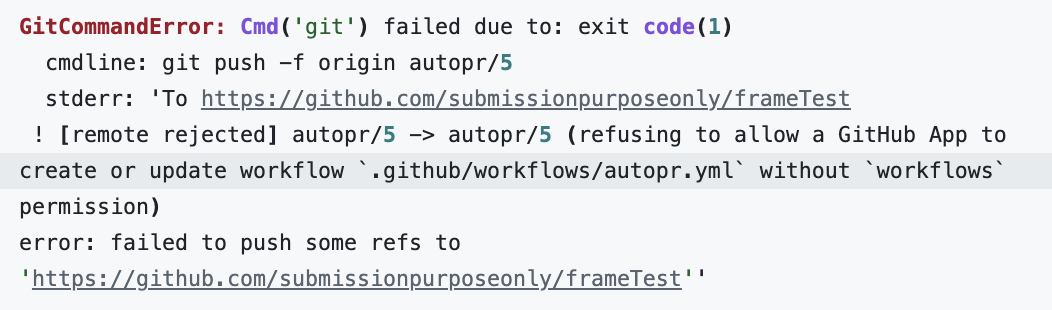}}
    \caption{The initial \texttt{ProtocolMismatchException} caught by SHIELDA during the Execution phase. The exception message explicitly states the lack of 'workflows' permission.}
    \label{fig:execution_failure_log}
\end{figure}

\textcolor{black}{The workflow begins as designed, with the agent ingesting the prompt from the GitHub issue. It successfully formulates a multi-commit plan which, unbeknownst to the agent, contains a prohibited action. The exception is raised at the end of the execution phase, when the agent attempts to push its local commits, including the modification to its own workflow file, to the remote repository. This action is rejected by the external system (GitHub), resulting in a runtime exception. SHIELDA's \texttt{Exception Classifier} immediately catches this runtime exception. Based on the output from the underlying Git command, it is initially classified as an execution-phase Exception, which we categorize as \texttt{ProtocolMismatchException}. The critical evidence from the failed run's log is shown in Figure \ref{fig:execution_failure_log}.}

\textcolor{black}{Following the SHIELDA framework, a simple local handling pattern (e.g., \texttt{P018: Retry with Backoff}) is first attempted by the \texttt{Handling Executor}. However, all retry attempts would fail identically. This exhaustion of the local handling strategy signals that a deeper diagnosis is required, triggering the next phase of the SHIELDA process.}

\subsubsection{Phase 2: Escalation and Cross-Phase Root Cause Analysis}
\begin{figure}[hbt!]
    \centering
    \fbox{\includegraphics[width=1\textwidth]{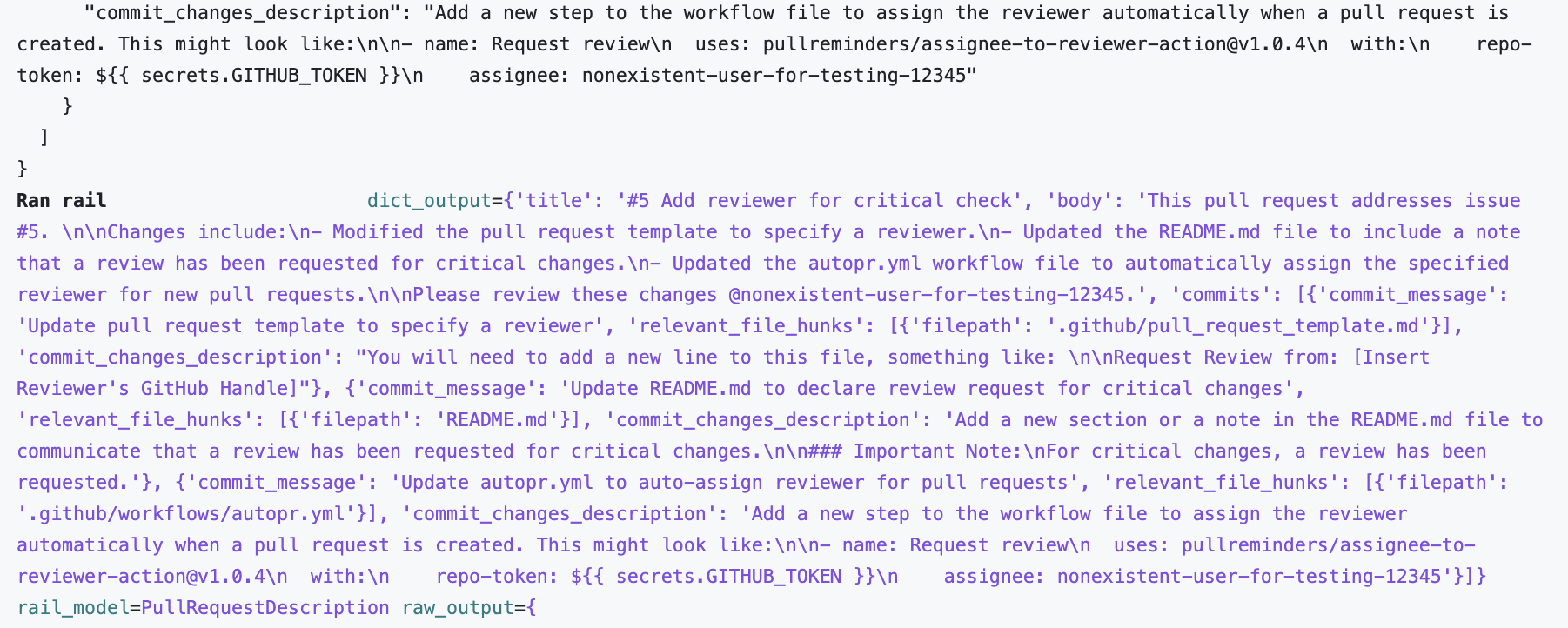}}
    \caption{The root cause of the exception, identified by SHIELDA. The agent's plan, generated during the Reasoning phase, explicitly contains the prohibited action.}
    \label{fig:reasoning_failure_evidence}
\end{figure}

When local handling fails, the process escalates to the \texttt{Escalation Controller} for a deeper, automated diagnosis. Employing a backward-chaining analysis of the \texttt{AgentOps} logs, the controller begins its investigation at the point of exception—the rejected git push command. It extracts the key entity from the error message, the file path .github/workflows/autopr.yml, and traces this artifact back through the workflow's history. This trace leads directly to the source of the exception in the planning phase logs: the original commit plan generated, which explicitly targeted the forbidden file (Figure~\ref{fig:reasoning_failure_evidence}).

\textcolor{black}{The \texttt{Escalation Controller} correctly re-classifies the problem. The root cause is not a faulty "git push" command, but a Faulty Task Structuring exception, where the agent created a plan that violates fundamental system constraints.}

\subsubsection{Phase 3: Plan Repair and Controlled Recovery}
\label{ssubsec:phase3_recovery}
\textcolor{black}{With the true root cause identified as a \texttt{Faulty Task Structuring} exception, SHIELDA's \texttt{Handler Pattern Registry} maps this exception to a handler pattern: \texttt{P012}. The \texttt{Handler Executor} then orchestrates the execution of this pattern's triadic mechanisms in a logical sequence to achieve a full recovery.}

\begin{figure}[hbt!]
    \centering
    \fbox{\includegraphics[width=1\textwidth]{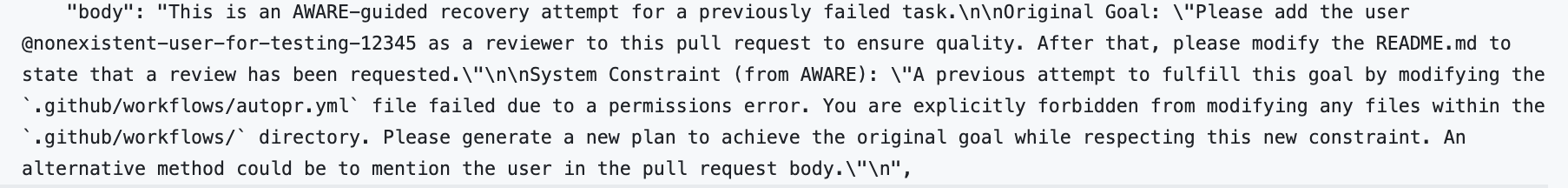}}
    \caption{The corrective prompt generated by SHIELDA's \texttt{Plan Repair} mechanism, which is then used to simulate the recovery run.}
    \label{fig:aware_corrected_prompt}
\end{figure}

\begin{figure}[hbt!]
    \centering
    \fbox{\includegraphics[width=1\textwidth]{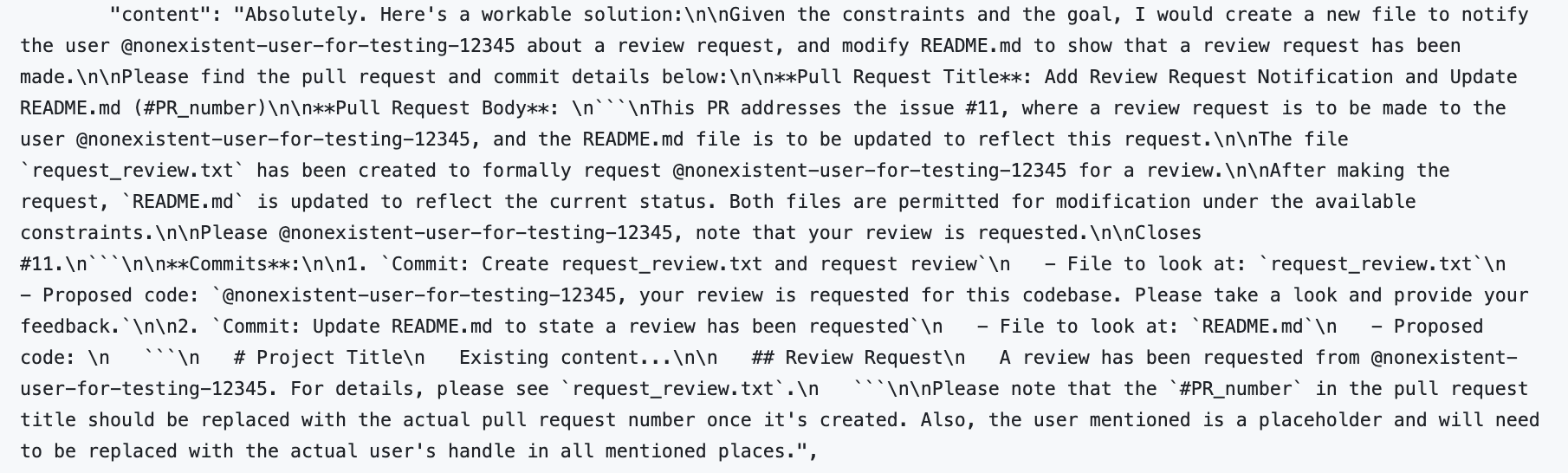}}
    \fbox{\includegraphics[width=1\textwidth]{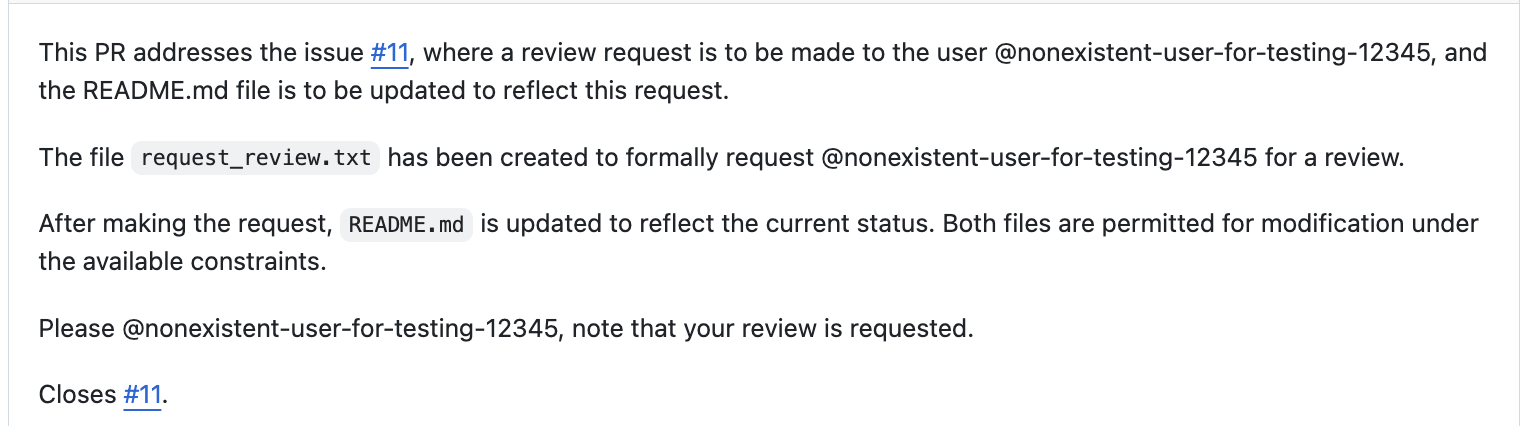}}
    \caption{The final successful outcome after SHIELDA-guided recovery. The agent has successfully completed the task by following the new, safe plan.}
    \label{fig:final_success}
\end{figure}

\textcolor{black}{The recovery process begins with the \texttt{Plan Repair} mechanism. To repair the flawed plan, SHIELDA's first action is to formulate a new, corrective directive for the \texttt{Reasoning Module}. It constructs a new prompt that includes both the original user goal and the new, explicit system constraint discovered during root cause analysis ("You are explicitly forbidden from modifying workflow files..."). This new prompt, shown in Figure \ref{fig:aware_corrected_prompt}, is designed to force the agent to generate a completely new, safe, and compliant plan. The agent's \texttt{Reasoning Module}, upon receiving this corrected input, successfully generates a new plan that avoids the prohibited action. The \texttt{Abort} mechanism is executed. Crucially, this does not terminate the agent's entire mission. Instead, it aborts the current, flawed execution thread that was operating on the faulty plan. This action gracefully stops any further operations based on the bad plan and clears the execution context, paving the way for a new, clean execution thread to begin based on the corrected plan obtained in the \texttt{Plan Repair} step. Finally, the \texttt{P012} pattern specifies a \texttt{No-op} (no operation) for state recovery. This is appropriate in this specific scenario because the failed git push command did not alter the state of the external system (the remote repository). While the agent's local workspace contains unpushed, faulty commits, they will be superseded by the new execution path. Thus, no complex compensation or external state rollback is required. As shown in Figure \ref{fig:final_success}. Guided by the new corrective directive generated during the \texttt{Plan Repair} step, the agent formulates a safe, alternative plan. It then successfully executes this new plan, creating a valid pull request that fulfills the original user goal without violating system constraints. This result serves as evidence that the SHIELDA-guided \texttt{P012} handler pattern is effective in correcting this class of reasoning-based failures.}

\subsection{Discussion}
\label{sec:discussion}
The successful resolution of the above case study provides compelling evidence supporting our primary research question (RQ3) concerning SHIELDA's runtime effectiveness. The execution trace detailed in Section \ref{sec:execution_trace} moves beyond a simple pass/fail metric, offering a nuanced view into the unique capabilities of our proposed exception handling framework. The findings highlight three critical aspects of the SHIELDA framework that address pervasive gaps in current exception handling approaches for LLM agents.

\subsubsection{Multi-stage, Phase-Aware Recovery}
This case study's primary finding is the validation of SHIELDA's ability to perform multi-stage, phase-aware recovery. The initial manifestation of failure was an \texttt{ExternalSystemPermissionException}, caught during the \textit{Execution} phase. A conventional, phase-disjoint handler would likely classify this as a non-recoverable infrastructure exception and terminate, as the immediate cause lies outside the agent's direct control. The true novelty of SHIELDA is showcased in its subsequent actions. After routine local handling (i.e., retry) fails, the \texttt{Escalation Controller} correctly infers that the symptom's origin may lie upstream. It successfully traces the exception back to a completely different domain and phase: a \texttt{Faulty Task Structuring} exception within the agent's \textit{Reasoning/Planning} phase. This ability to causally link a low-level system interaction exception to a high-level strategic planning flaw is a significant advancement. It demonstrates a holistic workflow understanding that is essential for resolving the complex, cascading exceptions common in autonomous agent systems.

\subsubsection{The Critical Role of Runtime Log Analysis}
The cross-phase exception handling would be impossible without treating logging data as a first-class runtime resource. Our case study illustrates that logs are not just for human debugging but are machine-readable inputs for the SHIELDA framework itself. The \texttt{Escalation Controller} leverages these structured logs to perform an automated trace, connecting key entities from a failed execution step (like a file path) back to the original plan object. This ability to turn logs into actionable intelligence is what enables SHIELDA to perform precise, automated root cause analysis.

\subsubsection{Closed-Loop Recovery via Constrained Re-planning}
Finally, this case study showcases SHIELDA's ability to perform exception handling in agentic workflow. After identifying the root cause, the framework does not simply terminate the workflow. Instead, the selected handler pattern (\texttt{P012}) leverages its \texttt{Plan Repair} capability. By injecting the system constraint back into the agent's context via a corrected prompt, SHIELDA fundamentally alters the problem space for the next reasoning cycle. This forces the agent to discard its initial, flawed plan and generate a new, safe alternative that respects the environment's permission boundaries. This complete "diagnose-repair-re-execute" cycle, where the framework actively corrects the agent's operational assumptions, presents a far more sophisticated and resilient paradigm than the linear, often terminal logic of traditional \texttt{try-catch} blocks, paving the way for more dependable agentic automation.

\section{Threats to Validity}
\label{sec:threats_to_validity}
\textcolor{black}{While our case study demonstrates the effectiveness of the SHIELDA framework's core recovery loop, we acknowledge several potential threats to the validity of our findings that warrant discussion.}

\subsection{Construct Validity}
\textcolor{black}{The primary threat to construct validity lies in our exception classification taxonomy. The framework's effectiveness is predicated on the \texttt{Exception Classifier}'s ability to map a runtime exception to one of the 36 predefined types in our taxonomy. The \texttt{ExternalSystemPermissionException} encountered in this case study, while fitting best under the \texttt{Protocol Mismatch} category, highlights that any real-world taxonomy may not be exhaustive. This underscores the necessity for the taxonomy to be an extensible, living artifact. While our framework is designed to support this, its performance on entirely novel or un-catalogued exception types that it cannot classify remains an open area for evaluation. A failure in this initial classification step could prevent the correct handler pattern from being invoked, thus undermining the entire recovery process. However, we have mitigated this threat by grounding our taxonomy in a systematic review to maximize its comprehensiveness. Furthermore, SHIELDA's architecture anticipates this contingency, as an unclassifiable exception can be routed to the \texttt{Escalation Controller} for fallback handling, which is a pathway for future work.}

\subsection{Internal Validity}
\textcolor{black}{A threat to internal validity stems from the engineered nature of the exception scenario. While the final case study was an authentic exception mode emerging from the agent's interaction with a real-world external system, the initial prompt was deliberately crafted to be ambiguous. This was designed to increase the likelihood that the agent would formulate the specific flawed plan we intended to study. This controlled approach allows for a reproducible evaluation of SHIELDA's recovery logic, but it may not fully represent the stochastic nature of reasoning exceptions that might occur from more benign user prompts. Future work should therefore involve a broader empirical study on naturally occurring exceptions to validate SHIELDA's effectiveness in more stochastic, real-world settings.} 

\subsection{External Validity}
\textcolor{black}{The primary threat to external validity is the generalizability of our findings from a single case study. While the SHIELDA framework itself is designed to be general-purpose, its evaluation in this paper is based on one agent and one specific exception pathway. The framework's performance on other types of agent architectures (e.g., multi-agent systems, agents with different toolsets) remains an open question for future work. Furthermore, another significant limitation is that the current implementation of the \texttt{Handler Pattern Registry} is static. It relies on pre-defined mappings from exception types to handling patterns. The framework does not currently feature "dynamic learning" to adapt or generate new handling strategies based on past outcomes. This may limit its effectiveness in novel situations where pre-defined patterns are suboptimal, a challenge that points toward future research in adaptive exception handling for LLM agents. We consider this static design as a necessary first step and a key direction for future work is to use the logged data from our framework to enable SHIELDA to learn and dynamically select the best handling pattern for a given exception.}

\section{Related Work}
\label{sec:related_work}
\subsection{Exception Handling in Traditional Software Engineering and Automation}
\textcolor{black}{Exception handling is a cornerstone of robust software engineering, with well-established techniques designed for deterministic, linear workflows. For decades, mechanisms such as \texttt{try-catch} blocks, system-level checkpointing, and manual failovers have served as the standard toolkit for addressing predictable errors like file access failures or network timeouts \cite{parnas1990evaluation, white2012hadoop}. In early AI, rule-based expert systems handled exceptions by targeting predefined conditions with static recovery actions, such as invoking a default rule or prompting for user intervention \cite{hayes1985rule}. This rule-based philosophy's modern heir is Robotic Process Automation (RPA), which deploys digital workers that diligently follow hard-coded scripts. The core weakness of this paradigm is its brittleness: these agents stumble at the slightest deviation from their prescribed path, such as a button changing location on a screen or a form field being renamed \cite{wornow2024automating}. As a result, the promise of RPA is often undermined by high initial setup costs, unreliable execution, and the burdensome need for constant human oversight and maintenance \cite{wornow2024automating}. Ultimately, both traditional methods and RPA systems are anchored in an assumption of deterministic exception modes. This foundation proves inadequate for the stochastic, dynamic, and semantically rich nature of agentic workflows.}

\subsection{Characterizing Exceptions in Agentic Workflows}
\textcolor{black}{The shift towards LLM-based agentic systems introduces a new spectrum of exceptions that go beyond traditional software errors. Research has begun to identify and categorize these failure modes, though often in a fragmented manner. A primary area of focus has been on the intrinsic limitations of the models themselves. This includes exceptions arising from flawed reasoning, such as producing contradictory logic \cite{sun2024ai}, getting stuck in circular reasoning loops \cite{yao2023react}, or generating factually incorrect statements, commonly known as hallucination \cite{zong2024triad}. When an agent acts upon such flawed reasoning, it inevitably leads to downstream execution failures. Another critical failure point is action validity and grounding. Studies have highlighted numerous execution-phase exceptions, including \textbf{Tool Invocation Exception} where an agent misuses an external tool \cite{xie2024travelplanner, milev2025toolfuzz, zhuo2025identifying}, and UI Element Misclick where it fails to interact with the correct graphical interface element \cite{xie2024osworld, bonatti2024windows}. Furthermore, research on improving exception handling within code itself has identified failures in the development process, such as "Insensitive Detection of Fragile Code" and "Inaccurate Capture of Exception Types" \cite{zhang2024seeker}. While these studies provide valuable insights into specific types of failures, a holistic understanding has been lacking.}

\subsection{Recovery and Mitigation Strategies for LLM Agents}
\textcolor{black}{In response to these newly identified exceptions, a variety of recovery and mitigation strategies have begun to emerge in the literature. Several efforts have focused on building specialized frameworks to tackle exceptions within narrow, high-stakes domains. For instance, RCAgent deploys autonomous agents with dedicated tools to perform Root Cause Analysis in complex cloud environments \cite{wang2023rcagent}. In the mobile domain, LLMPA automates multi-step app interactions and uses a ``Controllable Calibration'' module to validate predicted actions before execution \cite{guan2024intelligent}. Other approaches target the exception-handling process itself. HEALER, for example, proposes a dynamic self-healing system where an LLM generates and executes recovery code at runtime \cite{sun2024llm}, while SEEKER employs a multi-agent system to statically analyze and repair flawed exception handling logic in source code \cite{zhang2024seeker}.} \textcolor{black}{Concurrently, other research explores more fundamental techniques for agent resilience. A prominent theme is agent evolution through self-improvement, which includes methods like self-reflection, where an agent verbally reinforces its learning from task feedback \cite{shinn2023reflexion, chen2024survey}, and leveraging multi-agent conversations for collaborative error correction \cite{wu2023autogen, li2024survey}. While this body of work offers a rich set of recovery methods, they often exist as isolated techniques or are deeply embedded within monolithic, specialized systems. A unified, modular, and runtime-oriented framework that systematically organizes these strategies has been absent. Our work addresses these by shifting the focus from individual recovery strategies to a systematic, architectural approach for managing them.}

\section{Conclusion}
\label{sec:conclusion}
This paper addresses the critical challenge of managing diverse and unpredictable exceptions in LLM-based agentic workflows. We advocate for a shift from ad-hoc error mitigation to a more principled, engineering-based discipline for the systematic management of agent exceptions. First, we constructed a comprehensive taxonomy of 36 distinct exception types spanning both the reasoning, planning, and execution phases of agentic workflows. Second, building on this taxonomy, we designed and implemented SHIELDA, a composable, phase-aware runtime framework. Future work will proceed along two primary avenues: extending the framework to address complex multi-agent coordination exceptions, and integrating adaptive policy learning to dynamically optimize the selection of handling patterns.


\bibliographystyle{unsrt}
\bibliography{references}

\begin{thebibliography}{10}

\bibitem{mirzadeh2024gsm}
Iman Mirzadeh, Keivan Alizadeh, Hooman Shahrokhi, Oncel Tuzel, Samy Bengio, and Mehrdad Farajtabar.
\newblock Gsm-symbolic: Understanding the limitations of mathematical reasoning in large language models.
\newblock {\em arXiv preprint arXiv:2410.05229}, 2024.

\bibitem{xiong2023can}
Miao Xiong, Zhiyuan Hu, Xinyang Lu, Yifei Li, Jie Fu, Junxian He, and Bryan Hooi.
\newblock Can llms express their uncertainty? an empirical evaluation of confidence elicitation in llms.
\newblock {\em arXiv preprint arXiv:2306.13063}, 2023.

\bibitem{ruan2023tptu}
Jingqing Ruan, Yihong Chen, Bin Zhang, Zhiwei Xu, Tianpeng Bao, Hangyu Mao, Ziyue Li, Xingyu Zeng, Rui Zhao, et~al.
\newblock Tptu: Task planning and tool usage of large language model-based ai agents.
\newblock In {\em NeurIPS 2023 Foundation Models for Decision Making Workshop}, 2023.

\bibitem{chen2025towards}
Qiguang Chen, Libo Qin, Jinhao Liu, Dengyun Peng, Jiannan Guan, Peng Wang, Mengkang Hu, Yuhang Zhou, Te~Gao, and Wangxiang Che.
\newblock Towards reasoning era: A survey of long chain-of-thought for reasoning large language models.
\newblock {\em arXiv preprint arXiv:2503.09567}, 2025.

\bibitem{hou2025model}
Xinyi Hou, Yanjie Zhao, Shenao Wang, and Haoyu Wang.
\newblock Model context protocol (mcp): Landscape, security threats, and future research directions.
\newblock {\em arXiv preprint arXiv:2503.23278}, 2025.

\bibitem{jin2024llms}
Haolin Jin, Linghan Huang, Haipeng Cai, Jun Yan, Bo~Li, and Huaming Chen.
\newblock From llms to llm-based agents for software engineering: A survey of current, challenges and future.
\newblock {\em arXiv preprint arXiv:2408.02479}, 2024.

\bibitem{kong2024sharpunlockinginteractivehallucination}
Chuyi Kong, Ziyang Luo, Hongzhan Lin, Zhiyuan Fan, Yaxin Fan, Yuxi Sun, and Jing Ma.
\newblock Sharp: Unlocking interactive hallucination via stance transfer in role-playing agents, 2024.

\bibitem{zhang2025agent}
Shaokun Zhang, Ming Yin, Jieyu Zhang, Jiale Liu, Zhiguang Han, Jingyang Zhang, Beibin Li, Chi Wang, Huazheng Wang, Yiran Chen, et~al.
\newblock Which agent causes task failures and when? on automated failure attribution of llm multi-agent systems.
\newblock {\em arXiv preprint arXiv:2505.00212}, 2025.

\bibitem{chin2024human}
Daniel Chin, Yuxuan Wang, and Gus Xia.
\newblock Human-centered llm-agent user interface: A position paper.
\newblock {\em arXiv preprint arXiv:2405.13050}, 2024.

\bibitem{guan2024intelligent}
Yanchu Guan, Dong Wang, Zhixuan Chu, Shiyu Wang, Feiyue Ni, Ruihua Song, and Chenyi Zhuang.
\newblock Intelligent agents with llm-based process automation.
\newblock In {\em Proceedings of the 30th ACM SIGKDD Conference on Knowledge Discovery and Data Mining}, pages 5018--5027, 2024.

\bibitem{zhan2024injecagent}
Qiusi Zhan, Zhixiang Liang, Zifan Ying, and Daniel Kang.
\newblock Injecagent: Benchmarking indirect prompt injections in tool-integrated large language model agents.
\newblock {\em arXiv preprint arXiv:2403.02691}, 2024.

\bibitem{sun2024ai}
Yujie Sun, Dongfang Sheng, Zihan Zhou, and Yifei Wu.
\newblock Ai hallucination: towards a comprehensive classification of distorted information in artificial intelligence-generated content.
\newblock {\em Humanities and Social Sciences Communications}, 11(1):1--14, 2024.

\bibitem{yao2023react}
Shunyu Yao, Jeffrey Zhao, Dian Yu, Nan Du, Izhak Shafran, Karthik Narasimhan, and Yuan Cao.
\newblock React: Synergizing reasoning and acting in language models.
\newblock In {\em International Conference on Learning Representations (ICLR)}, 2023.

\bibitem{xie2024travelplanner}
Jian Xie, Kai Zhang, Jiangjie Chen, Tinghui Zhu, Renze Lou, Yuandong Tian, Yanghua Xiao, and Yu~Su.
\newblock Travelplanner: A benchmark for real-world planning with language agents.
\newblock {\em arXiv preprint arXiv:2402.01622}, 2024.

\bibitem{chen2025agentpoison}
Zhaorun Chen, Zhen Xiang, Chaowei Xiao, Dawn Song, and Bo~Li.
\newblock Agentpoison: Red-teaming llm agents via poisoning memory or knowledge bases.
\newblock {\em Advances in Neural Information Processing Systems}, 37:130185--130213, 2025.

\bibitem{hatalis2023memory}
Kostas Hatalis, Despina Christou, Joshua Myers, Steven Jones, Keith Lambert, Adam Amos-Binks, Zohreh Dannenhauer, and Dustin Dannenhauer.
\newblock Memory matters: The need to improve long-term memory in llm-agents.
\newblock In {\em Proceedings of the AAAI Symposium Series}, volume~2, pages 277--280, 2023.

\bibitem{salama2025meminsightautonomousmemoryaugmentation}
Rana Salama, Jason Cai, Michelle Yuan, Anna Currey, Monica Sunkara, Yi~Zhang, and Yassine Benajiba.
\newblock Meminsight: Autonomous memory augmentation for llm agents, 2025.

\bibitem{zong2024triad}
Chang Zong, Yuchen Yan, Weiming Lu, Jian Shao, Eliot Huang, Heng Chang, and Yueting Zhuang.
\newblock Triad: A framework leveraging a multi-role llm-based agent to solve knowledge base question answering.
\newblock {\em arXiv preprint arXiv:2402.14320}, 2024.

\bibitem{microsoft2025taxonomy}
{Microsoft AI Red Team}.
\newblock New whitepaper outlines the taxonomy of failure modes in ai agents.
\newblock \url{https://www.microsoft.com/en-us/security/blog/2025/04/24/new-whitepaper-outlines-the-taxonomy-of-failure-modes-in-ai-agents/}, April 2025.
\newblock Accessed: 2025-05-09.

\bibitem{xu2024knowledge}
Rongwu Xu, Zehan Qi, Zhijiang Guo, Cunxiang Wang, Hongru Wang, Yue Zhang, and Wei Xu.
\newblock Knowledge conflicts for llms: A survey.
\newblock {\em arXiv preprint arXiv:2403.08319}, 2024.

\bibitem{liu2024lost}
Nelson~F Liu, Kevin Lin, John Hewitt, Ashwin Paranjape, Michele Bevilacqua, Fabio Petroni, and Percy Liang.
\newblock Lost in the middle: How language models use long contexts.
\newblock {\em Transactions of the Association for Computational Linguistics}, 12:157--173, 2024.

\bibitem{owasp2025output}
{OWASP Foundation}.
\newblock Owasp top 10 for llm applications 2025.
\newblock \url{https://genai.owasp.org}, 2025.
\newblock LLM05: Improper Output Handling.

\bibitem{sun2024toolsfaildetectingsilent}
Jimin Sun, So~Yeon Min, Yingshan Chang, and Yonatan Bisk.
\newblock Tools fail: Detecting silent errors in faulty tools, 2024.

\bibitem{zhang2025logiagent}
Ke~Zhang, Chenxi Zhang, Chong Wang, Chi Zhang, YaChen Wu, Zhenchang Xing, Yang Liu, Qingshan Li, and Xin Peng.
\newblock Logiagent: Automated logical testing for rest systems with llm-based multi-agents.
\newblock {\em arXiv preprint arXiv:2503.15079}, 2025.

\bibitem{milev2025toolfuzz}
Ivan Milev, Mislav Balunovi{\'c}, Maximilian Baader, and Martin Vechev.
\newblock Toolfuzz--automated agent tool testing.
\newblock {\em arXiv preprint arXiv:2503.04479}, 2025.

\bibitem{xie2024osworld}
Tianbao Xie, Danyang Zhang, Jixuan Chen, Xiaochuan Li, Siheng Zhao, Ruisheng Cao, Toh~Jing Hua, Zhoujun Cheng, Dongchan Shin, Fangyu Lei, et~al.
\newblock Osworld: Benchmarking multimodal agents for open-ended tasks in real computer environments.
\newblock {\em arXiv preprint arXiv:2404.07972}, 2024.

\bibitem{bonatti2024windows}
Rogerio Bonatti, Dan Zhao, Francesco Bonacci, Dillon Dupont, Sara Abdali, Yinheng Li, Yadong Lu, Justin Wagle, Kazuhito Koishida, Arthur Bucker, et~al.
\newblock Windows agent arena: Evaluating multi-modal os agents at scale.
\newblock {\em arXiv preprint arXiv:2409.08264}, 2024.

\bibitem{cemri2025multi}
Mert Cemri, Melissa~Z Pan, Shuyi Yang, Lakshya~A Agrawal, Bhavya Chopra, Rishabh Tiwari, Kurt Keutzer, Aditya Parameswaran, Dan Klein, Kannan Ramchandran, et~al.
\newblock Why do multi-agent llm systems fail?
\newblock {\em arXiv preprint arXiv:2503.13657}, 2025.

\bibitem{lu2024responsiblegenerativeaireference}
Qinghua Lu, Liming Zhu, Xiwei Xu, Zhenchang Xing, Stefan Harrer, and Jon Whittle.
\newblock Towards responsible generative ai: A reference architecture for designing foundation model based agents, 2024.

\bibitem{dong2024taxonomy}
Liming Dong, Qinghua Lu, and Liming Zhu.
\newblock A taxonomy of agentops for enabling observability of foundation model based agents.
\newblock {\em arXiv preprint arXiv:2411.05285}, 2024.

\bibitem{lee2024promptinfection}
Donghyun Lee and Mo~Tiwari.
\newblock Prompt infection: Llm-to-llm prompt injection within multi-agent systems.
\newblock {\em arXiv preprint arXiv:2410.07283}, 2024.

\bibitem{kutasov2025shade}
Jonathan Kutasov, Yuqi Sun, Paul Colognese, Teun van~der Weij, Linda Petrini, Chen Bo~Calvin Zhang, John Hughes, Xiang Deng, Henry Sleight, Tyler Tracy, et~al.
\newblock Shade-arena: Evaluating sabotage and monitoring in llm agents.
\newblock {\em arXiv preprint arXiv:2506.15740}, 2025.

\bibitem{yuan2025agent}
Siyu Yuan, Zehui Chen, Zhiheng Xi, Junjie Ye, Zhengyin Du, and Jiecao Chen.
\newblock Agent-r: Training language model agents to reflect via iterative self-training.
\newblock {\em arXiv preprint arXiv:2501.11425}, 2025.

\bibitem{wu2024avatar}
Shirley Wu, Shiyu Zhao, Qian Huang, Kexin Huang, Michihiro Yasunaga, Kaidi Cao, Vassilis Ioannidis, Karthik Subbian, Jure Leskovec, and James~Y Zou.
\newblock Avatar: Optimizing llm agents for tool usage via contrastive reasoning.
\newblock {\em Advances in Neural Information Processing Systems}, 37:25981--26010, 2024.

\bibitem{zhang2024survey}
Zeyu Zhang, Xiaohe Bo, Chen Ma, Rui Li, Xu~Chen, Quanyu Dai, Jieming Zhu, Zhenhua Dong, and Ji-Rong Wen.
\newblock A survey on the memory mechanism of large language model based agents.
\newblock {\em arXiv preprint arXiv:2404.13501}, 2024.

\bibitem{xiong2025memorymanagementimpactsllm}
Zidi Xiong, Yuping Lin, Wenya Xie, Pengfei He, Jiliang Tang, Himabindu Lakkaraju, and Zhen Xiang.
\newblock How memory management impacts llm agents: An empirical study of experience-following behavior, 2025.

\bibitem{luo2025oneke}
Yujie Luo, Xiangyuan Ru, Kangwei Liu, Lin Yuan, Mengshu Sun, Ningyu Zhang, Lei Liang, Zhiqiang Zhang, Jun Zhou, Lanning Wei, et~al.
\newblock Oneke: A dockerized schema-guided llm agent-based knowledge extraction system.
\newblock In {\em Companion Proceedings of the ACM on Web Conference 2025}, pages 2871--2874, 2025.

\bibitem{zhang2024knowhalu}
Jiawei Zhang, Chejian Xu, Yu~Gai, Freddy Lecue, Dawn Song, and Bo~Li.
\newblock Knowhalu: Hallucination detection via multi-form knowledge based factual checking.
\newblock {\em arXiv preprint arXiv:2404.02935}, 2024.

\bibitem{winstontaxonomy}
Cailin Winston and Ren{\'e} Just.
\newblock A taxonomy of failures in tool-augmented llms.

\bibitem{wang2024rethinking}
Qineng Wang, Zihao Wang, Ying Su, Hanghang Tong, and Yangqiu Song.
\newblock Rethinking the bounds of llm reasoning: Are multi-agent discussions the key?
\newblock {\em arXiv preprint arXiv:2402.18272}, 2024.

\bibitem{zhuo2025identifying}
Terry~Yue Zhuo, Junda He, Jiamou Sun, Zhenchang Xing, David Lo, John Grundy, and Xiaoning Du.
\newblock Identifying and mitigating api misuse in large language models.
\newblock {\em arXiv preprint arXiv:2503.22821}, 2025.

\bibitem{shi2025tool}
Zhengliang Shi, Shen Gao, Lingyong Yan, Yue Feng, Xiuyi Chen, Zhumin Chen, Dawei Yin, Suzan Verberne, and Zhaochun Ren.
\newblock Tool learning in the wild: Empowering language models as automatic tool agents.
\newblock In {\em Proceedings of the ACM on Web Conference 2025}, pages 2222--2237, 2025.

\bibitem{gim2024asynchronous}
In~Gim, Seung-seob Lee, and Lin Zhong.
\newblock Asynchronous llm function calling.
\newblock {\em arXiv preprint arXiv:2412.07017}, 2024.

\bibitem{russell2006workflow}
Nick Russell, Wil~MP van~der Aalst, and Arthur~HM ter Hofstede.
\newblock Workflow exception patterns.
\newblock In {\em International Conference on Advanced Information Systems Engineering}, pages 288--302. Springer, 2006.

\bibitem{parnas1990evaluation}
David~L Parnas, A~John Van~Schouwen, and Shu~Po Kwan.
\newblock Evaluation of safety-critical software.
\newblock {\em Communications of the ACM}, 33(6):636--648, 1990.

\bibitem{white2012hadoop}
Tom White.
\newblock {\em Hadoop: The definitive guide}.
\newblock " O'Reilly Media, Inc.", 2012.

\bibitem{hayes1985rule}
Frederick Hayes-Roth.
\newblock Rule-based systems.
\newblock {\em Communications of the ACM}, 28(9):921--932, 1985.

\bibitem{wornow2024automating}
Michael Wornow, Avanika Narayan, Krista Opsahl-Ong, Quinn McIntyre, Nigam~H Shah, and Christopher R{\'e}.
\newblock Automating the enterprise with foundation models.
\newblock {\em arXiv preprint arXiv:2405.03710}, 2024.

\bibitem{zhang2024seeker}
Xuanming Zhang, Yuxuan Chen, Yuan Yuan, and Minlie Huang.
\newblock Seeker: Enhancing exception handling in code with llm-based multi-agent approach.
\newblock {\em arXiv preprint arXiv:2410.06949}, 2024.

\bibitem{wang2023rcagent}
Zefan Wang, Zichuan Liu, Yingying Zhang, Aoxiao Zhong, Lunting Fan, Lingfei Wu, and Qingsong Wen.
\newblock Rcagent: Cloud root cause analysis by autonomous agents with tool-augmented large language models.
\newblock {\em arXiv preprint arXiv:2310.16340}, 2023.

\bibitem{sun2024llm}
Zhensu Sun, Haotian Zhu, Bowen Xu, Xiaoning Du, Li~Li, and David Lo.
\newblock Llm as runtime error handler: A promising pathway to adaptive self-healing of software systems.
\newblock {\em arXiv preprint arXiv:2408.01055}, 2024.

\bibitem{shinn2023reflexion}
Noah Shinn, Federico Cassano, Ashwin Gopinath, Karthik Narasimhan, and Shunyu Yao.
\newblock Reflexion: Language agents with verbal reinforcement learning.
\newblock {\em Advances in Neural Information Processing Systems}, 36:8634--8652, 2023.

\bibitem{chen2024survey}
Shuaihang Chen, Yuanxing Liu, Wei Han, Weinan Zhang, and Ting Liu.
\newblock A survey on llm-based multi-agent system: Recent advances and new frontiers in application.
\newblock {\em arXiv preprint arXiv:2412.17481}, 2024.

\bibitem{wu2023autogen}
Qingyun Wu, Gagan Bansal, Jieyu Zhang, Yiran Wu, Shaokun Zhang, Erkang Zhu, Beibin Li, Li~Jiang, Xiaoyun Zhang, and Chi Wang.
\newblock Autogen: Enabling next-gen llm applications via multi-agent conversation framework.
\newblock {\em arXiv preprint arXiv:2308.08155}, 2023.

\bibitem{li2024survey}
Xinzhe Li.
\newblock A survey on llm-based agents: Common workflows and reusable llm-profiled components.
\newblock {\em arXiv preprint arXiv:2406.05804}, 2024.

\end{thebibliography}

\newpage
\appendix
\section{Tables}
\label{sec:appendix_tables}
\begin{table}[h]
\caption{Local Handling Mechanisms.}
\label{tab:local_handling}
\centering
\scriptsize
\begin{tabular}{p{1cm} p{3.5cm} p{8cm}}
\toprule
\textbf{\#} & \textbf{Mechanism} & \textbf{Description} \\
\midrule
1  & Clarify Prompt            & Ask the user to clarify ambiguous or underspecified goals. \\
2  & Echo Validation           & Paraphrase and reflect the goal to verify understanding. \\
3  & Context Tagging           & Label context content to separate user/system input sources. \\
4  & Default Interpretation    & Apply default meaning when input is unclear. \\
5  & Disentangled Prompting    & Separate memory/context/KB input paths explicitly. \\
6  & Prompt Rewriting          & Rewrite or optimize the original prompt. \\
7  & Prompt Sanitization       & Clean injections, redundancy, or special characters from prompt. \\
8  & Graph Validation          & Check the logical structure of the reasoning chain. \\
9  & KB Trust Scoring          & Score knowledge by trust level, filter untrusted entries. \\
10 & Logic Re-ranking          & Rerank multiple reasoning paths. \\
11 & Recursive Checkpointing   & Insert intermediate checks to avoid infinite loops. \\
12 & Abort Task Chain          & Abort when the plan is unrecoverable. \\
13 & Conflict Resolution       & Resolve agent conflicts via rule or confirmation. \\
14 & Constraint Pruning        & Remove conflicting or unfulfillable planning constraints. \\
15 & Forward Chaining          & Predict if a task is feasible before continuing. \\
16 & Peer Confirmation         & Confirm cross-agent collaboration. \\
17 & Plan Repair               & Repair broken or incomplete plan structures. \\
18 & Plan Shortening           & Shorten long plan sequences. \\
19 & Role-based Check          & Ensure the agent is capable of the assigned role. \\
20 & Subgoal Reordering        & Reorder subtasks to improve execution logic. \\
21 & Attribute Filtering       & Remove incorrect features during memory matching. \\
22 & Escalate UI Failure       & Escalate UI failure, uncertainty to humans. \\
23 & External Call Timeout     & Fallback handler when the external system/API times out. \\
24 & Fallback                  & Output default template if structure is broken. \\
25 & Fallback to Alternate API & Try another API endpoint if the primary fails. \\
26 & Low-confidence Filter     & Skip or downgrade model outputs with low confidence. \\
27 & Memory Slot Isolation     & Isolate faulty memory slots to prevent propagation. \\
28 & Oracle Verification       & Use rules/verifiers to judge model/tool correctness. \\
29 & Output Sanitization       & Clean unsafe or injected segments in generated output. \\
30 & Output Truncation         & Truncate model outputs that exceed token limits. \\
31 & Protocol Downgrade        & Use a more compatible older protocol if needed. \\
32 & Reset Memory              & Clear or rollback corrupted memory. \\
33 & Response Normalization    & Normalize API output to standard format. \\
34 & Retry with Backoff        & Retry tool/API call with exponential delay. \\
35 & Sampling Adjustment       & Adjust decoding parameters like temperature/top-p. \\
36 & Schema Validation         & Check if output matches the required schema. \\
37 & Semantic Constraint Checking & Check output against semantic constraints or rules. \\
38 & Switch Tool               & Replace failed tool/API with a backup. \\
39 & Timeout Escalation        & Trigger escalation on tool/API timeout. \\
40 & Escalate to Human         & General fallback strategy, escalate to human when needed. \\

\bottomrule
\end{tabular}
\end{table}

\begin{table}[h]
\caption{Flow Control Mechanisms.}
\label{tab:flow_control}
\centering
\scriptsize
\begin{tabular}{p{1cm} p{3.5cm} p{8cm}}
\toprule
\textbf{\#} & \textbf{Mechanism} & \textbf{Description} \\
\midrule
1 & Continue & The agent proceeds to the next step after successful local recovery. \\
2 & Skip & The agent bypasses the current subgoal or step. \\
3 & Abort & The agent terminates the current task or execution thread. \\
\bottomrule
\end{tabular}
\end{table}

\begin{table}[h]
\caption{State Recovery Mechanisms.}
\label{tab:state_recovery}
\centering
\scriptsize
\begin{tabular}{p{1cm} p{3.5cm} p{8cm}}
\toprule
\textbf{\#} & \textbf{Mechanism} & \textbf{Description} \\
\midrule
1 & No-op & No state repair is needed; the exception has no persistent side effects. \\
2 & Rollback & Restore the agent's internal state (e.g., memory) to a clean checkpoint. \\
3 & Compensate & Trigger a logical reversal to undo externally executed side effects. \\
\bottomrule
\end{tabular}
\end{table}

\begin{table*}[ht]
\renewcommand{\arraystretch}{1.1}
\caption{Exception Handling Pattern Table}
\label{tab:pattern_combination_full_fixed}
\centering
\scriptsize
\begin{tabular}{p{1.5cm} p{4.5cm} p{3cm} p{3cm}}
\toprule
\textbf{Pattern ID} & \textbf{Local Handling} & \textbf{Flow Control} & \textbf{State Recovery} \\
\midrule
P001 & Clarify Prompt & Abort & No-op \\
P002 & Clarify Prompt & Continue & No-op \\
P003 & Echo Validation & Continue & No-op \\
P004 & Prompt Rewriting & Abort & No-op \\
P005 & Prompt Sanitization & Continue & No-op \\
P006 & Context Tagging & Continue & No-op \\
P007 & Default Interpretation & Continue & No-op \\
P008 & Graph Validation & Abort & No-op \\
P009 & Recursive Checkpointing & Abort & No-op \\
P010 & Logic Re-ranking & Continue & No-op \\
P011 & KB Trust Scoring & Continue & No-op \\
P012 & Plan Repair & Abort & No-op \\
P013 & Plan Shortening & Abort & No-op \\
P014 & Forward Checking & Abort & No-op \\
P015 & Constraint Pruning & Abort & No-op \\
P016 & Subgoal Reordering & Abort & No-op \\
P017 & Abort Task Chain & Abort & Rollback \\
P018 & Retry with Backoff & Continue & No-op \\
P019 & Switch Tool & Abort & No-op \\
P020 & Fallback to Alternate API & Abort & No-op \\
P021 & Schema Validation & Abort & No-op \\
P022 & Schema Validation & Continue & No-op \\
P023 & Semantic Constraint Checking & Continue & No-op \\
P024 & Oracle Verification & Continue & Rollback \\
P025 & Output Sanitization & Continue & No-op \\
P026 & Fallback Template & Continue & No-op \\
P027 & Reset Memory & Abort & Rollback \\
P028 & Memory Slot Isolation & Continue & No-op \\
P029 & Attribute Filtering & Continue & Rollback \\
P030 & Disentangled Prompting & Abort & No-op \\
P031 & Response Normalization & Continue & No-op \\
P032 & Timeout Escalation & Skip & Compensate \\
P033 & Timeout Escalation & Abort & Compensate \\
P034 & Low-confidence Filter & Continue & No-op \\
P035 & Output Truncation & Continue & No-op \\
P036 & Sampling Adjustment & Continue & No-op \\
P037 & Escalate to Human & Skip & Compensate \\
P038 & Escalate to Human & Skip & No-op \\
P039 & Escalate to Human & Abort & Compensate \\
P040 & Escalate to Human & Abort & No-op \\
P041 & Protocol Downgrade & Abort & No-op \\
P042 & External Call Timeout Fallback & Skip & Compensate \\
P043 & External Call Timeout Fallback & Skip & No-op \\
P044 & External Call Timeout Fallback & Abort & Compensate \\
P045 & External Call Timeout Fallback & Abort & No-op \\
P046 & Role-based Check & Abort & No-op \\
P047 & Conflict Resolution Prompt & Abort & No-op \\
P048 & Peer Confirmation & Continue & No-op \\
\bottomrule
\end{tabular}
\end{table*}

\begin{table*}[ht]
\renewcommand{\arraystretch}{1.2}
\caption{Exception to Primary Handler Pattern Mapping: Representative Examples}
\label{tab:exception_to_pattern_hero}
\centering
\scriptsize
\begin{tabularx}{\linewidth}{p{2.8cm} p{1.6cm} p{1.2cm} p{2.2cm} p{1.8cm} p{1.8cm}}
\toprule
\textbf{Exception Type} & \textbf{Artifact} & \textbf{Pattern ID} & \textbf{Local Handling} & \textbf{Flow Control} & \textbf{State Recovery} \\
\midrule
Ambiguous Goal & Goal & P001 & Clarify Prompt & Abort & No-op \\
\midrule
Contradictory Reasoning & Reasoning & P008 & Graph Validation & Abort & No-op \\
Faulty Task Structuring & Planning & P012 & Plan Repair & Abort & No-op \\
\midrule
Memory Poisoning & Memory & P027 & Reset Memory & Abort & Rollback \\
Hallucinated Facts & Knowledge Base & P011 & KB Trust Scoring & Continue & No-op \\
\midrule
Tool Invocation Exception & Tool & P018 & Retry with Backoff & Continue & No-op \\
Tool Output Exception & Tool & P021 & Schema Validation & Abort & No-op \\
API Invocation Exception & Interface & P018 & Retry with Backoff & Continue & No-op \\
UI Element Misclick & Interface & P037 & Escalate to Human & Skip & Compensate \\
\midrule
Error Propagation & Task Flow & P017 & Abort Task Chain & Abort & Rollback \\
Agent Conflict & Other Agent & P047 & Conflict Resolution Prompt & Abort & No-op \\
Protocol Mismatch & External System & P041 & Protocol Downgrade & Abort & No-op \\
\bottomrule
\end{tabularx}
\end{table*}







\end{document}